\documentclass[useAMS,usenatbib]{mn2e/mn2e}
\usepackage{graphicx}
\usepackage{txfonts}
\usepackage{hyperref}
\usepackage[usenames]{color}
\newcommand {\reac}[6] {$\rm\,{}^{#2}\kern-0.8pt{#1}\,({#3}\,,{#4})  \,{}^{#6}\kern-0.8pt{#5}\,$}
\newcommand{\object}{}

\newcommand{\Msun}{\mbox{${\rm~M_\odot}\,$}}

\newcommand{\sub}[1]{\mbox{$_{\rm #1}$}}

\newcommand{\Teff}{\mbox{$T\sub{eff}$}}

\newcommand{\logL}{\mbox{$\log(L/L_{\odot})$}}
\newcommand{\HP}{\mbox{$H_{\rm{P}}\,$}}

\newcommand{\beq}{\begin{equation}}
\newcommand{\eeq}{\end{equation}}
\newcommand{\beqa}{\begin{eqnarray}}
\newcommand{\eeqa}{\end{eqnarray}}
\newcommand{\benu}{\begin{enumerate}}
\newcommand{\eenu}{\end{enumerate}}
\newcommand{\bite}{\begin{itemize}}
\newcommand{\eite}{\end{itemize}}
\newcommand{\bdes}{\begin{description}}
\newcommand{\edes}{\end{description}}

\newcommand{\comment}[1]{}
\title[Low Metallicity Massive Stars with \sl{PARSEC}]{New {\sl PARSEC} evolutionary tracks of massive stars at low metallicity:
testing canonical stellar evolution in nearby star forming dwarf  galaxies.}
\author{}

\author[Tang et al.]{Jing Tang$^1$,  Alessandro Bressan$^1$,
Philip Rosenfield$^2$, Alessandra Slemer$^2$, Paola Marigo$^2$,
\and L\'eo Girardi$^3$,  Luciana Bianchi$^4$\\
  $^1$ SISSA, via Bonomea 265, I-34136 Trieste, Italy \\
  $^2$ Dipartimento di Fisica e Astronomia Galileo Galilei,
  Universit\`a di Padova, Vicolo dell'Osservatorio 3, I-35122 Padova, Italy \\
  $^3$ Osservatorio Astronomico di Padova, Vicolo dell'Osservatorio 5,
  I-35122 Padova, Italy \\
  $^4$ Dept of Physics and Astronomy, The Johns Hopkins University,
  407 Bloomberg center, 3400 N. Charles st., Baltimore, MD 21218, USA  \\
}

\begin{document}
\date{Accepted 2014 September 26.  Received 2014 September 25; in original form 2014 August 8}

\pagerange{\pageref{firstpage}--\pageref{lastpage}} \pubyear{2014}

\maketitle

\label{firstpage}

\begin{abstract}
We extend the {\sl\,PARSEC} library of stellar evolutionary tracks by computing new models of massive stars, from 14\Msun to 350\Msun. The input physics is the same used in the {\sl\,PARSEC}~V1.1 version, but for the mass-loss rate which is included by considering the most recent updates in literature.
We focus on
 low metallicity, $Z$=0.001 and $Z$=0.004, for which the metal poor dwarf
 irregular star forming galaxies, Sextans A, WLM and NCG6822, provide
 simple but powerful workbenches. The models reproduce fairly  well
 the observed CMDs but the
stellar colour distributions indicate that the predicted blue loop is
not hot enough
 in models with canonical extent of overshooting.
 In the framework of a mild extended mixing during central hydrogen
 burning, the only way
to reconcile the discrepancy
is to enhance the overshooting at the base of the convective envelope (EO) during
the first dredge-UP.
The mixing scales required to reproduce the observed loops, EO=2\HP
or EO=4\HP, are definitely larger than those
 derived from, e.g., the observed location of the RGB bump in low mass
 stars.
 This effect, if confirmed, would  imply a
strong dependence of the mixing scale below the formal Schwarzschild
border, on the stellar
mass or luminosity. Reproducing  the features of the observed CMDs with
standard values   of envelope overshooting would require a
  metallicity significantly lower than the values measured in these
  galaxies. Other quantities,
 such as the star formation rate and the initial mass function, are
 only slightly sensitive to this effect.
Future investigations will
consider other metallicities
 and different mixing schemes.
\end{abstract}


\section{Introduction}
The PADOVA database of stellar evolutionary tracks and isochrones has
been continuously updated during the last decades
\citep{Girardi_etal02,Marigo_etal08,Bertelli_etal09},
except for the models of the most massive stars ($M>$20\Msun)
that have remained untouched for two decades, after \citet{Bertelli_etal94}. 
In the most recent major revision, all the basic input physics was
revisited and updated, and
additions 
were introduced, such as the pre-main-sequence phase and very low mass
stellar models (0.35\Msun$\geq{M}\geq$0.1\Msun).  The 
 {\sl\,PARSEC} {\sl\,P}adova {T\sl{R}}ieste {\sl\,S}tellar
 {\sl\,E}volution {\sl\,C}ode is 
described in detail in \citet{Bressan_etal12,Bressan_etal13} and \citet{Chen_etal14}.
The PADOVA database is one of the most widely used sources of stellar evolutionary tracks and the {\sl\,PARSEC} revision is not complete without including models of the most massive stars.
With the aim of completing the library of stellar evolutionary tracks and isochrones computed with {\sl\,PARSEC}, we present the calculations of new evolutionary tracks of massive stars, from 14\Msun to 350\Msun
\footnote{The new updated and homogeneous sets of evolutionary tracks, from very low ($M$=0.1\Msun) to very massive ($M$=350\Msun) stars, are available at
http://stev.oapd.inaf.it}.
In this paper, the first of a series devoted to the evolution of
massive stars using {\sl\,PARSEC}, we deal with low metallicity environments.
We present the new
evolutionary tracks, and we
perform preliminary comparisons with
 observed colour magnitude diagrams (CMD). These tests are fundamental to validate
the new models. Models with
other values of
 metallicity are being calculated
and will be presented in future works.

Understanding the evolution of massive stars at low metallicity is
particularly relevant, as these stars drive the chemical and
dynamical evolution of the surrounding medium, and are a key
ingredient in modelling galaxy evolution at early times
\citep{Groh_etal13, Woosley_heger12}.
An advantage of studying massive stars at low metallicity is also that,
in the very local Universe, star formation is sparkling in gas-rich
galaxies which are also metal poor.
In the last  decades, the high spatial resolution of the space-borne HST, and the
collecting power of 8m telescopes from
the ground have provided accurate photometry of individual stars in
these galaxies, and metallicity estimates of young stars and gas
from spectroscopy.  Such data enable comparison of
observed CMDs with model
simulations at known metallicity,
an important test of how the canonical theory of
evolution of massive and intermediate mass stars performs at low metallicity.

The internal evolution of massive stars
is affected by two complex physical phenomena
for which there is still a significant lack of knowledge.
The first is the mass-loss rate, which becomes important in the domain
of the very massive stars (VMS; initial mass $M\gtrsim$30\Msun
depending on
metallicity). The second is internal mixing, usually from either
differential rotation or convective overshooting.
Both effects play an important role in stellar evolution because they may significantly modify the structure of the stars in an irreversible way and so their further evolution.
Among the observational quantities that are most sensitive to internal mixing are the predicted surface abundance ratios of CNO elements \citep{martins_palacios_13,Bouret-etal13}.
In  massive  stars the surface abundances can be affected by mixing
during the hydrogen burning phases and by the eventual subsequent
first dredge-up. However, the comparison of model predictions with
observed surface abundances still shows
significant discrepancies \citep{maeder_et_al14}.
Another interesting effect of mixing in this mass interval is the
location  in the CMD
 and duration of the ``blue loops'',
where the stars spend a fraction of their
central helium burning life-time.
The morphology and star population of
the Blue and Red Helium Burning Sequences (BHeBS and RHeBS) may depend on several other parameters, such as
the $\mathrm{^{12}C(\alpha,\gamma)^{16}O}$ reaction rate
\citep{Iben66,Bertelli_etal85,Brunish_etal90}, and
the  $\mathrm{^{14}N(p,\gamma)^{15}O}$ reaction rate  \citep{Xu_Li04,Weiss_etal05,Halabi_etal12},
but the most important parameters are the efficiency of mixing
in the convective core \citep{Bertelli_etal85} and
below the convective envelope  \citep{Alongi_etal91,Godart13}.

While the analysis of surface abundances provides a detailed view
of individual stars, the analysis of the stellar distribution across a CMD provides a complementary view of the duration of evolutionary phases over a broad range of masses.
The latter requires complete, well studied and populous stellar samples.
In this paper we use published photometry of three
star-forming dwarf galaxies in the Local Group,  that have hundreds of
thousands
resolved individual stars, complete  down to intermediate mass stars,
to build the observed CMDs for comparison with our new models.
In section \ref{sec:tracks} we describe the new stellar evolutionary tracks of massive stars calculated with {\sl\,PARSEC}. Apart from the inclusion of
mass loss, which had not been considered in stars less massive than $M=$12\Msun,
the other parameters are the standard ones in the published  tracks
\citep{Bressan_etal12}. However, having in mind the comparison with
observed CMDs, we present here also models with enhanced envelope overshooting,
which is known to favour more extended blue loops during central He burning.
In section \ref{sec:data} we briefly describe the  comparison data used,
and estimate the contamination by foreground
Milky Way stars. This is significant only in NGC6822, which is at low Galactic latitude.
The adopted CMD simulation technique
is outlined in
section \ref{sec:scmd}.
 Attenuation by interstellar (IS) dust is accounted for  using results from
the multi-band photometry by \citet{Bianchi_etal12}.
The comparison of the canonical model predictions with the observed CMDs of three selected galaxies, Sextans~A, WLM and NGC~6822, is discussed in section
\ref{sec:results} while those based on models with enhanced envelope overshooting
are shown in section \ref{simhrenvov}.
The main conclusions drawn from these comparisons are
outlined in section \ref{sec:conclusion}.
\section{New evolutionary tracks of massive stars with {\sl\,PARSEC}}
\label{sec:tracks}
The  largest value of initial mass in the previously published  set of evolutionary tracks computed with {\sl\,PARSEC}~V1.1 is $M$=12\Msun. The corresponding
Hydrogen burning lifetime is
about 15${\rm Myr}$. 
 To simulate
younger and most massive stars we usually resort to our previous database \citep{Bressan_etal93A&AS_100_647B,Fagotto_etal94A&AS104_365F,Fagotto_etal94A&AS105_29F,
Fagotto_etal94A&AS105_39F}. Since
our stellar evolution code has been deeply revised recently 
\citep{Girardi_etal00,Bressan_etal12,Bressan_etal13},
we present here 
new sets of evolutionary tracks of massive stars computed with the 
{\sl\,PARSEC} code. The masses range from $M$=14\Msun to $M$=350\Msun and 
 the evolution is followed
from the pre-main sequence phase to the
beginning of the carbon-burning phase.  
Together with the already published stellar evolution tracks 
the database now includes updated and homogeneous
sets of evolutionary tracks from very low ($M$=0.1\Msun) to very massive ($M$=350\Msun) stars, from the pre-main sequence to the beginning of central carbon burning.  
Here we describe the models with
 low metallicity, $Z$=0.001 to $Z$=0.004, that will be used in the following sections
to build
simulated CMDs for comparison
with those of three dwarf irregualr galaxies of
similar metallicity. Models for lower and higher
metallicity  will be presented and discussed in a forthcoming paper.
\begin{figure}
\includegraphics[angle=0,width=0.42\textwidth]{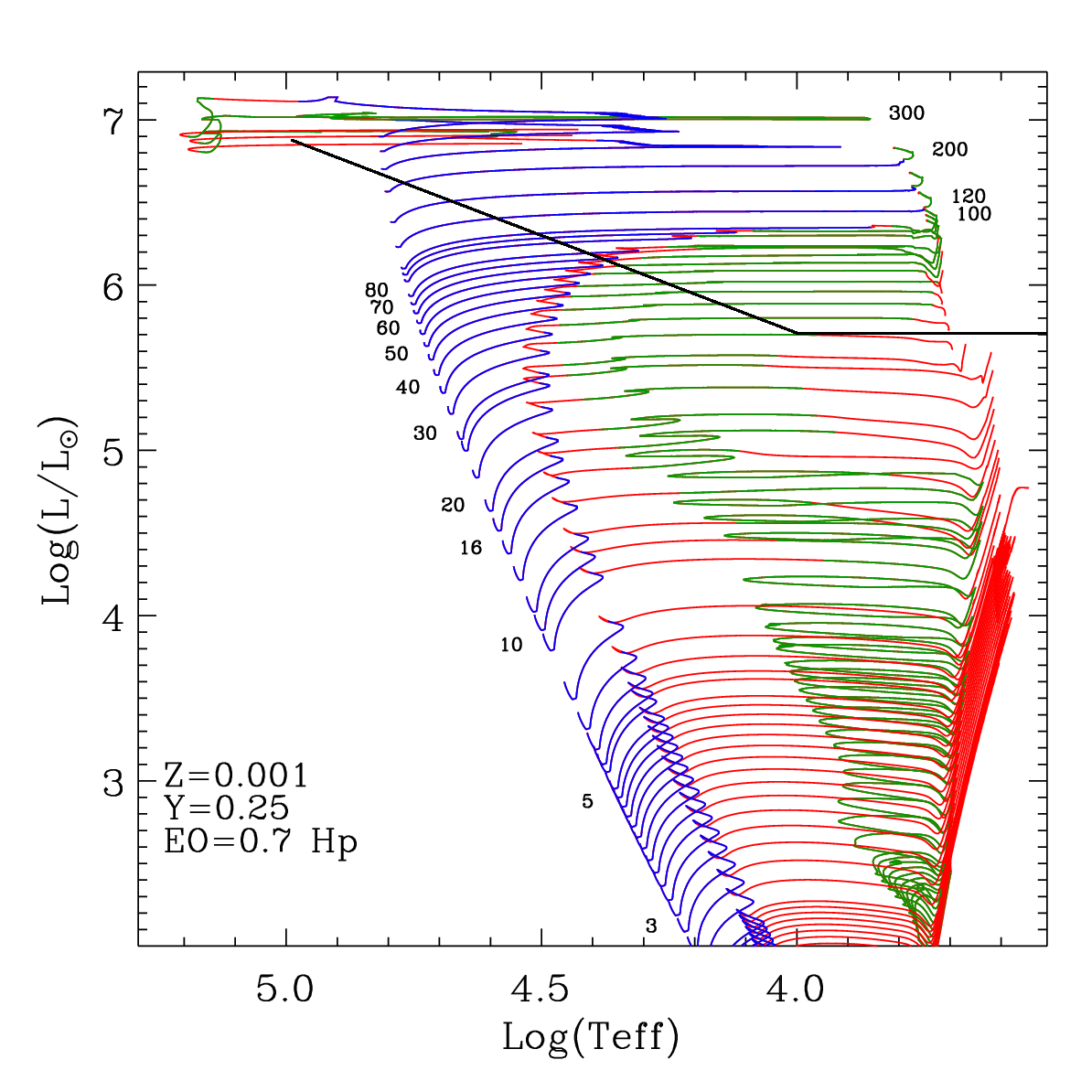}
\includegraphics[angle=0,width=0.42\textwidth]{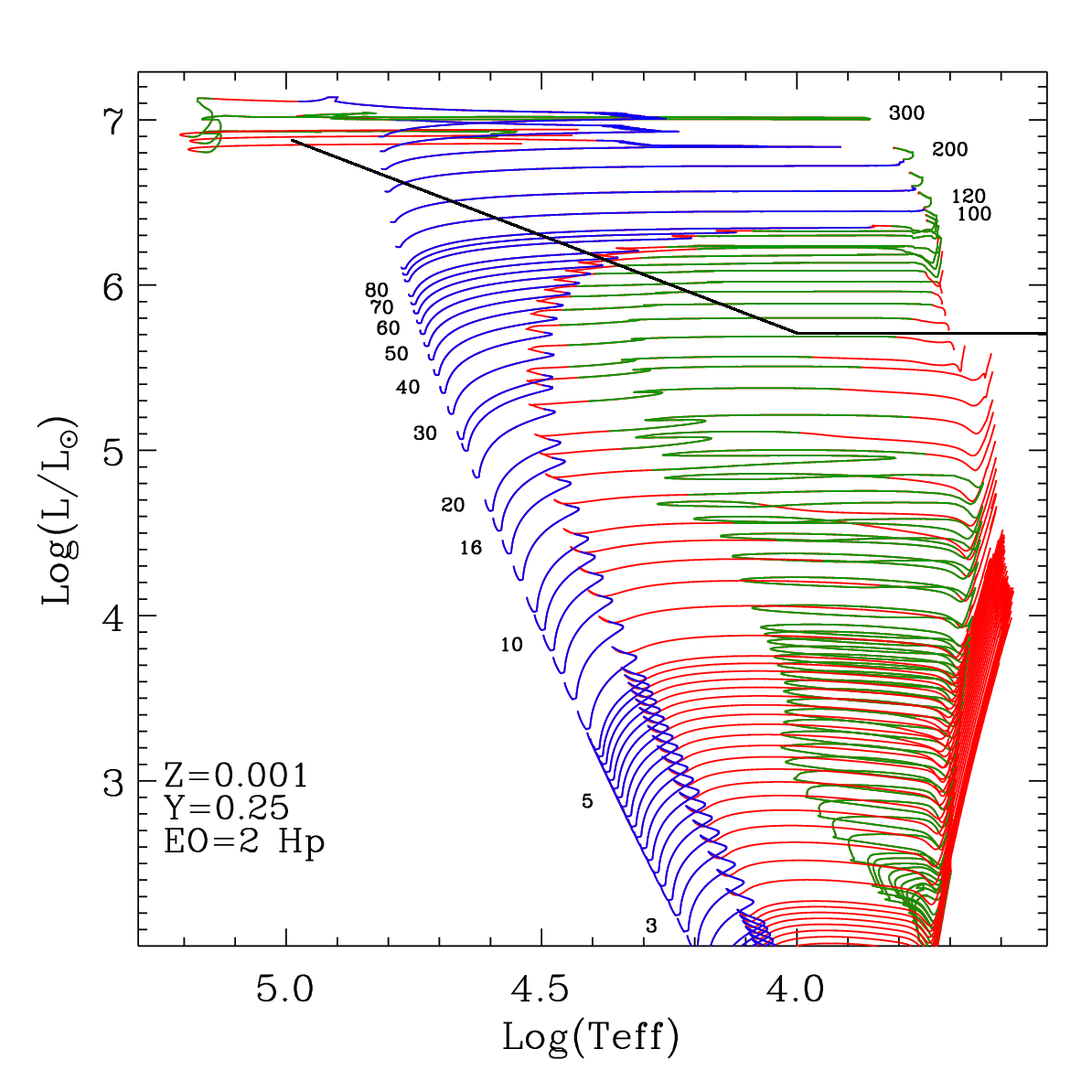}
\includegraphics[angle=0,width=0.42\textwidth]{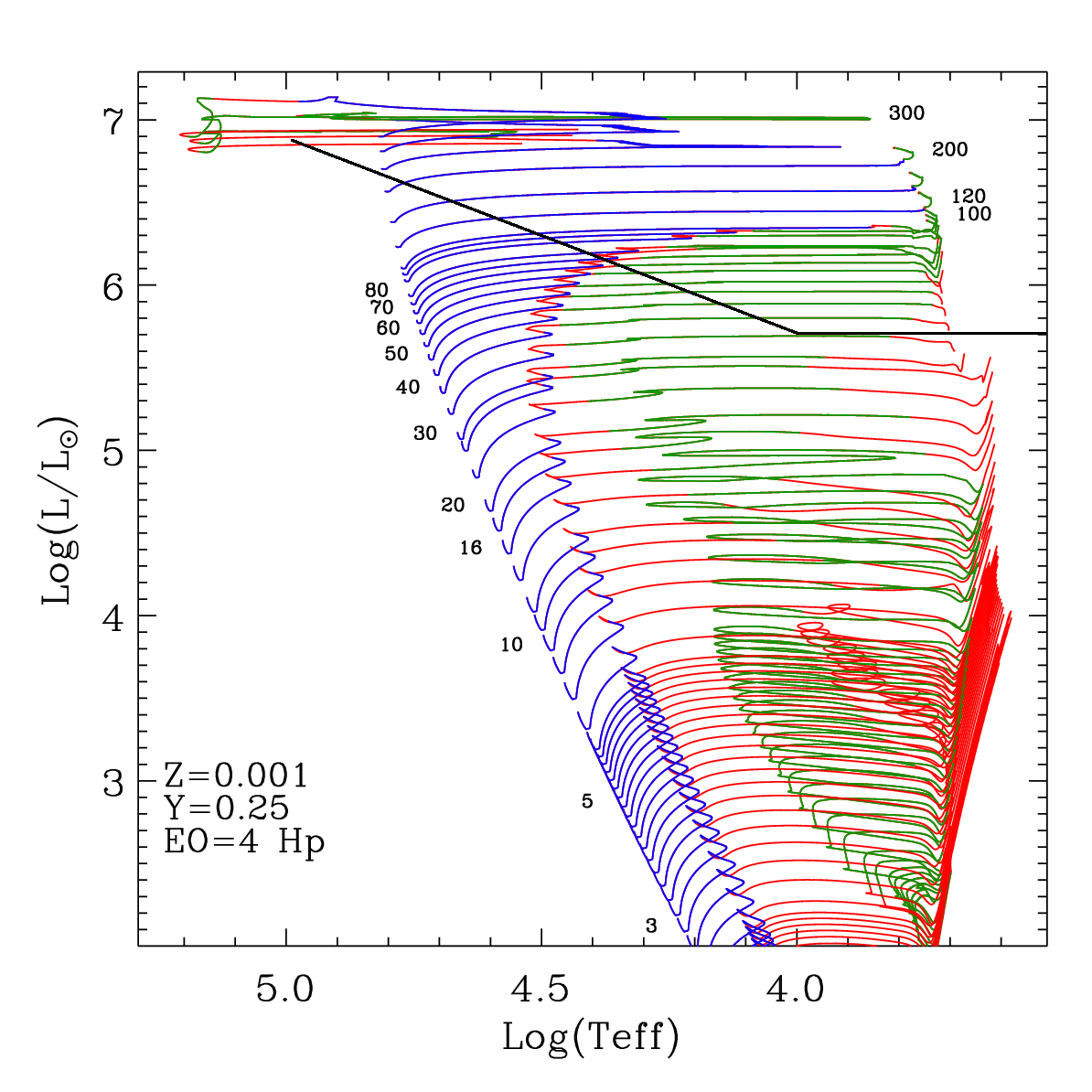}
\caption{New {\sl\,PARSEC} evolutionary tracks of massive stars
together with already published tracks of intermediate mass stars
for $Z$=0.001. The three panels refer to the different values of
envelope overshooting (EO) adopted, as indicated in the labels.
The black
lines indicates the observed \citet{Humphreys_Davidson_1979} transition limit at solar metallicity.
\label{fig_HRD_EO_Z001}}
\end{figure}
\begin{figure}
\includegraphics[angle=0,width=0.42\textwidth]{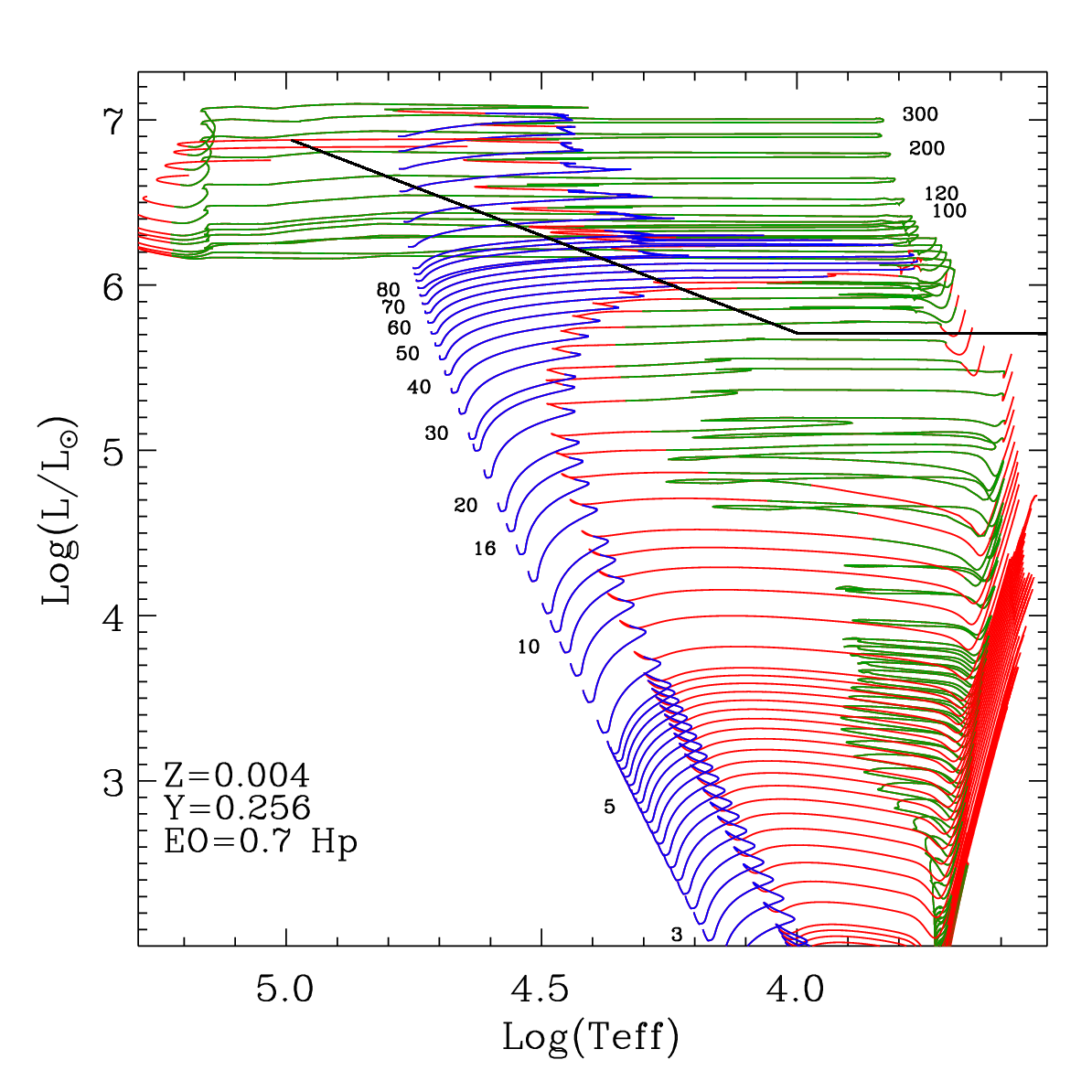}
\includegraphics[angle=0,width=0.42\textwidth]{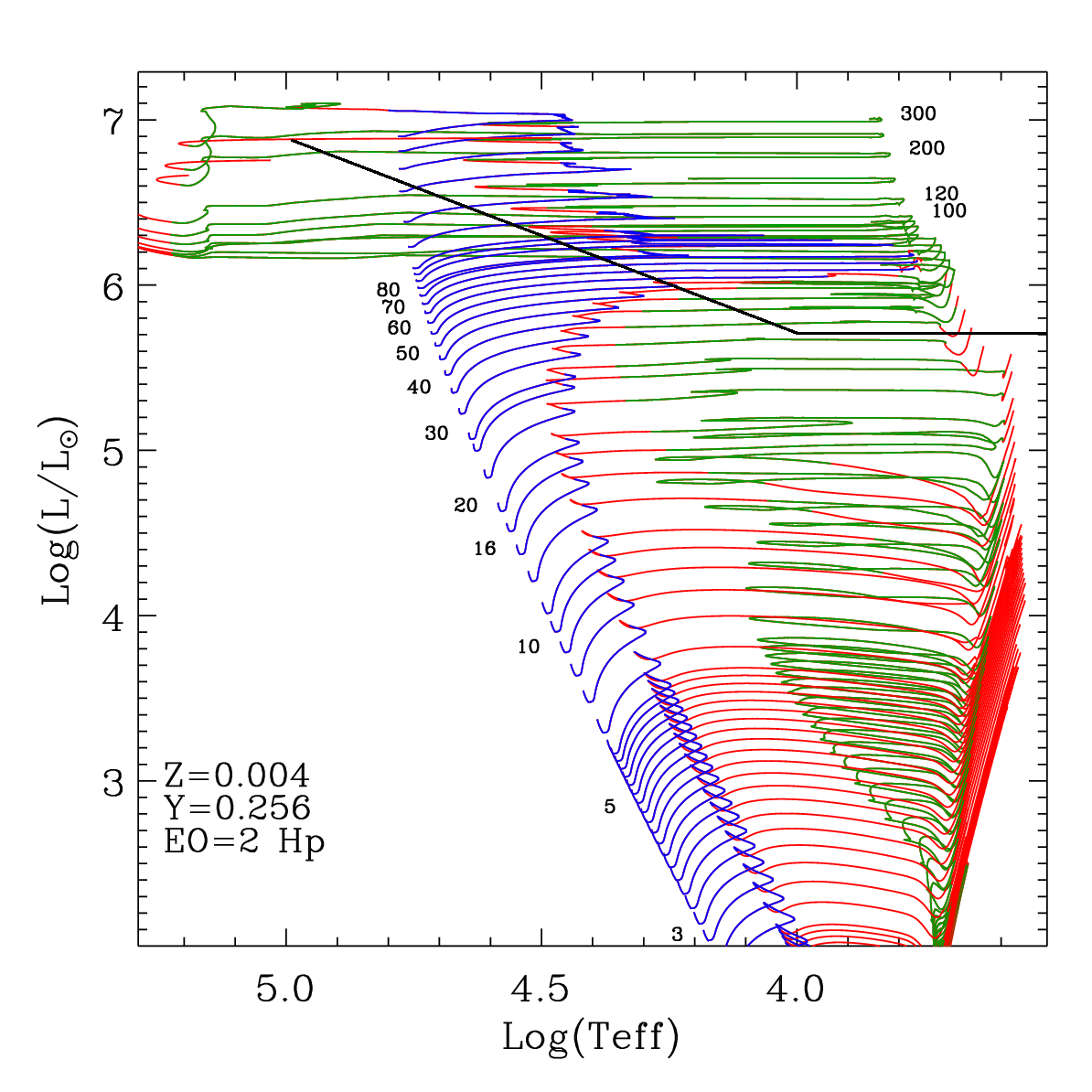}
\includegraphics[angle=0,width=0.42\textwidth]{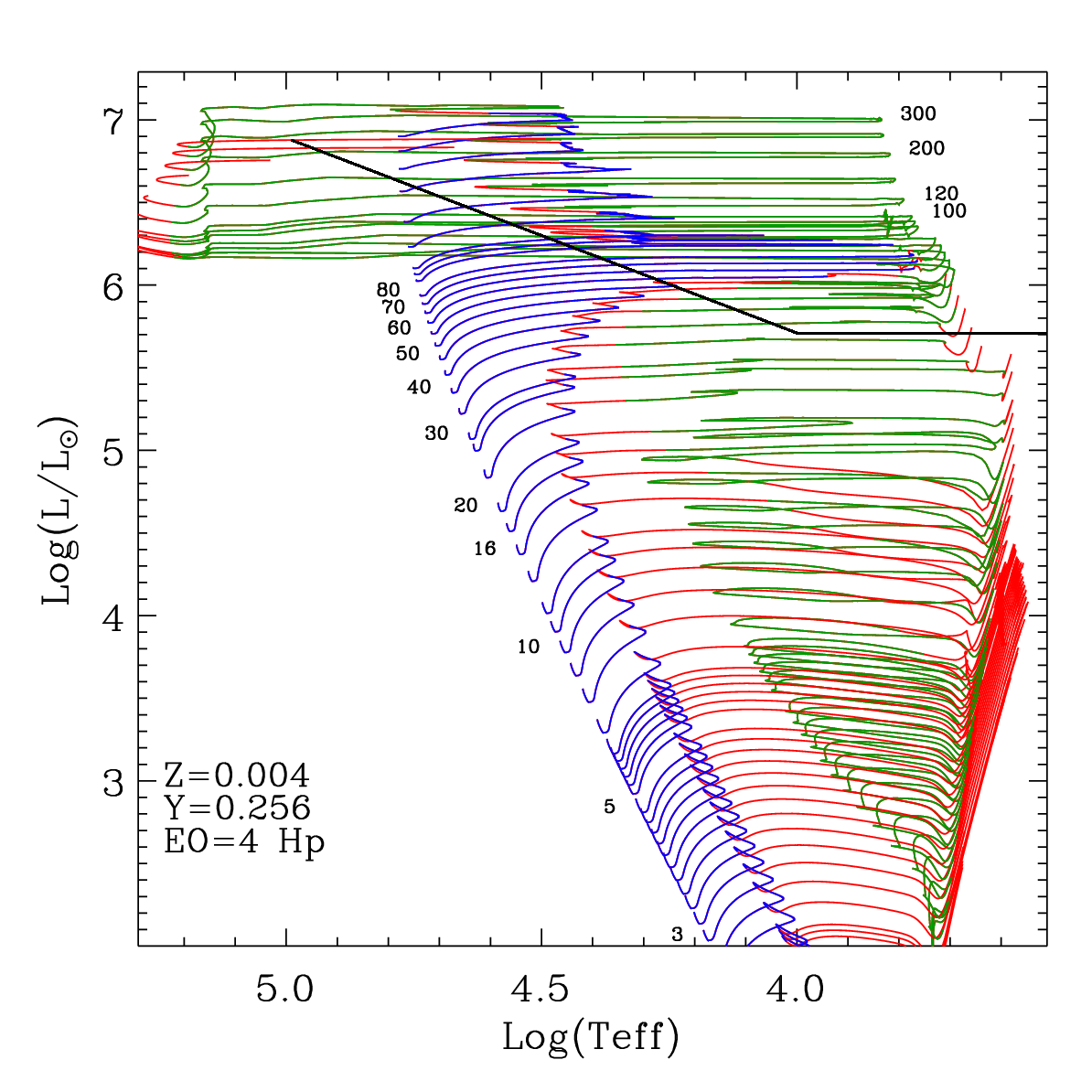}
\caption{Same as in Figure \ref{fig_HRD_EO_Z001} but for $Z$=0.004. As in the previous figure, the mass range shown is between about 3\Msun and 350\Msun. The colour codes blue, green and red, highlight the central H-burning, central He-burning and expansion/contraction phases, respectively. A few tracks are labelled with the initial mass.\label{fig_HRD_EO_Z004}}
\end{figure}

\subsection{Basic physics input}
The {\sl\,PARSEC} code is extensively discussed  in \citet{Bressan_etal12, Bressan_etal13} and  \citet{Chen_etal14};
here we briefly describe the most important updates.
The equation of state (EOS) is computed with the FreeEOS code
(A.W.~Irwin, \footnote{http://freeeos.sourceforge.net/}).
Opacities in the high-temperature regime, $4.2 \le \log(T/{\rm K}) \le 8.7$, are obtained from the Opacity Project At Livermoren(OPAL) 
\citep{IglesiasRogers_96},
while in the low-temperature regime, 
$3.2 \le \log(T/{\rm K}) \le 4.1$, we use opacities generated with our
\AE SOPUS\footnote{http://stev.oapd.inaf.it/aesopus} code \citep{MarigoAringer_09}.
Conductive opacities are included following \citet{Itoh_etal08}.
The nuclear reaction network consists of the p-p chains, the CNO tri-cycle, the Ne--Na and Mg--Al chains and the most important $\alpha$-capture reactions, including the $\alpha$-n reactions.
The reaction rates and the corresponding $Q$-values
are taken from the recommended rates in the JINA  database
\citep{Cyburt_etal10}.
Electron screening factors for all reactions are from
\citet{Dewitt_etal73} and \citet{Graboske_etal73}.
Energy losses by electron neutrinos are taken from
\citet{Munakata_etal85} and \citet{ItohKohyama_83} and \citet{Haft_etal94}.
The outer boundary conditions are discussed in \citet{Chen_etal14}.
The reference solar partition of heavy elements
is taken from \citet{Caffau_etal11} who revised
a few species of the \citet{GrevesseSauval_98} compilation (GS98).
According to \citet{Caffau_etal11}, the present-day Sun's
metallicity is $Z_{\odot}= 0.01524$, intermediate between the
most recent estimates, e.g.  $Z_{\odot}= 0.0141$ of \citet{Lodders_etal09} or $Z_{\odot}= 0.0134$ of \citet{Asplund_etal09} (AGSS),
and the previous value of $Z_{\odot}= 0.017$ by GS98.
\subsection{Convection and mixing}
The convective energy transport is described according
to the mixing-length theory of \citet{mlt},
with the mixing length parameter $\alpha_{\rm MLT}$
calibrated on the solar model, $\alpha_{\rm MLT}=1.74$.
The location of the boundary of the convective core of massive stars is estimated according to \citet{Bressan_etal81}
taking into account overshooting from the central
unstable region.  As thoroughly described in \citet{Bressan_etal13}, 
the main parameter describing core overshooting
is the mean free path of convective elements {\em across} the border of the unstable region,
expressed
in units of pressure scale height, $l_{\rm c}$=$\Lambda_{\rm c}$\HP.
The  overshooting parameter used in the core,
$\Lambda_{\rm c}=0.5$, is the result of the calibration obtained by the analysis of intermediate age clusters \citep{Girardi_etal09}
as well as individual stars \citep{Kamath_etal10, Deheuvels_etal10, Torres_etal14}.
Note 
 that the overshooting region obtained with this formalism
extends over about  0.5$l_{\rm c}$ 
{\em above} the unstable region.
Instability against convection is tested by means of the Schwarzschild criterion.
An alternative criterion  is  that of  Ledoux, which takes into account
also the effects of the gradient in the mean molecular weight.
Without overshooting, using the Ledoux criterion instead of the
Schwarzschild one leads to significant differences in the structure
of the models and in their evolution, mainly because the size of the
semi-convective regions above the unstable core depends on the adopted
criterion \citep{chiosi_summa_70}.  On the other hand, in presence of a
sizable convective overshooting the difference between models computed with the two different criteria
is significantly reduced.
In the present work we opt for the Schwarzschild criterion, which
is more appropriate to account for the effects of thermal dissipation,
according to \citet{Kato66}.

Overshooting at the base of the convective envelope (EO) can also
give rise to
sizable extra-mixing.
In the past, two important observational effects
have been related to this phenomenon, the location of the bump
in the red giant branch of low mass stars
(RGB Bump seen in the CMD of
globular clusters and old open clusters) and the
extent of the blue loops of intermediate mass stars \citep{Alongi_etal91}.
Both   features
 have been found to be   best explained by a moderate amount of overshooting
{\em below the border of the unstable region},
of about EO=0.25-1.0, in units of \HP.
More recently a similar mechanism
has been invoked to improve the agreement with the physical
state of matter in the convective-radiative transition region derived from solar
oscillation data \citep{Christensen-Dalsgaard_etal11}.
The standard value for
 intermediate and massive stars used in {\sl\,PARSEC}~V1.1 is EO=0.7\HP.
However,
as we will see later,
with this value it turns out to be difficult to reproduce the extended blue loops seen in the
metal-poor dwarf galaxies, therefore we have computed
two additional sets   of models increasing the $EO$
to match the observed  extended blue loops.
We adopted EO=2\HP and EO=4\HP.
These additional sets will be discussed in section \ref{hrenvov}.
\begin{figure*}
\includegraphics[angle=0,width=0.3\textwidth]{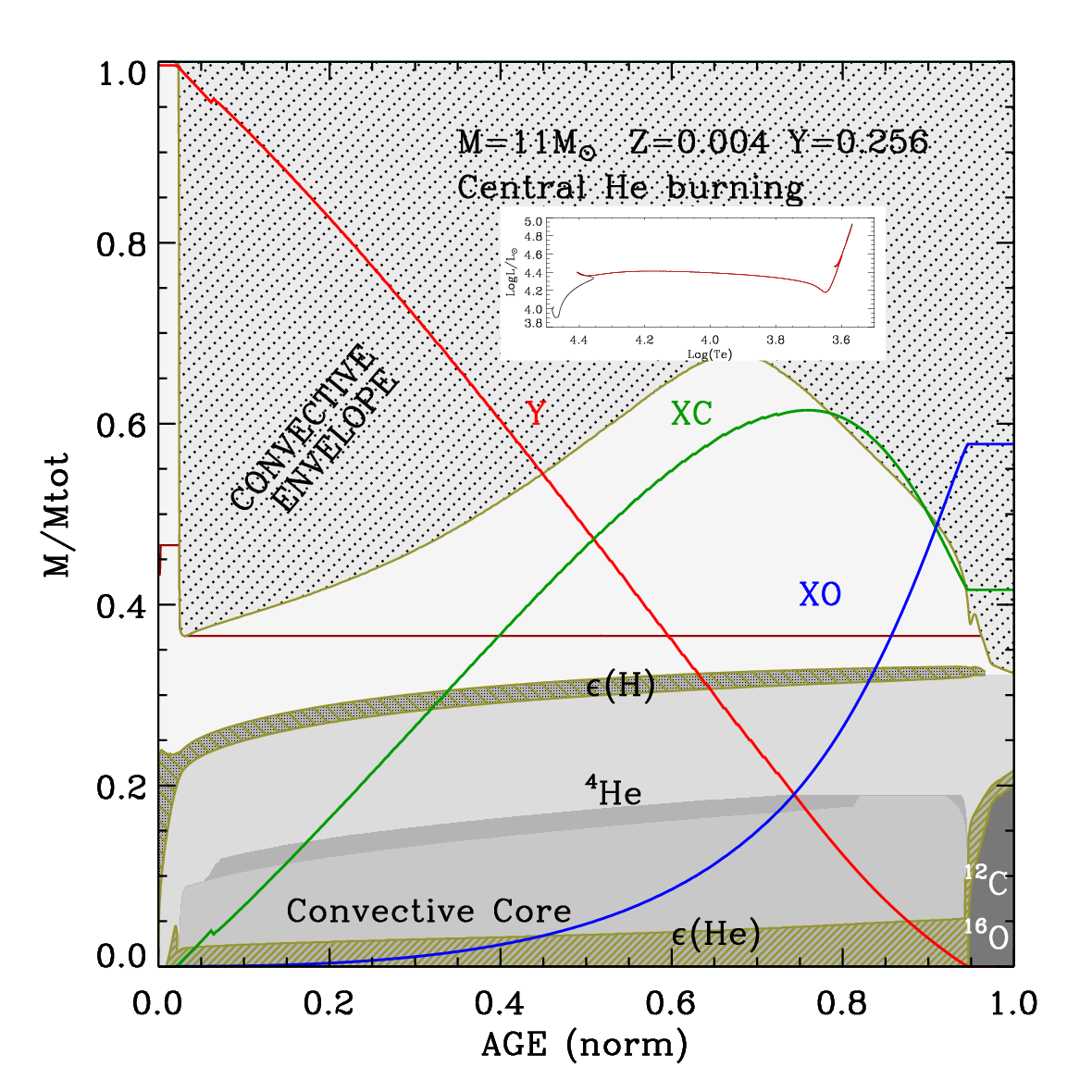}
\includegraphics[angle=0,width=0.3\textwidth]{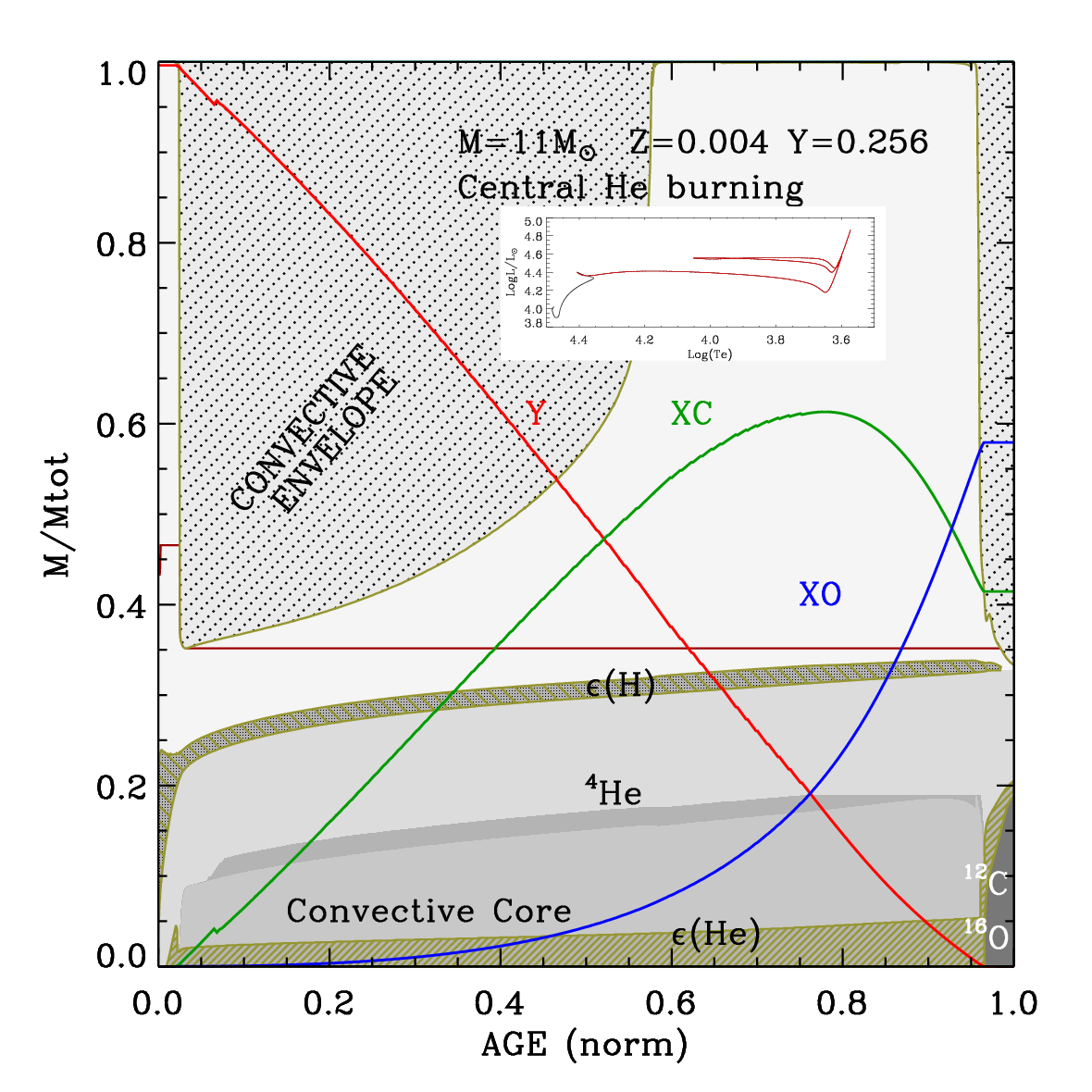}
\includegraphics[angle=0,width=0.3\textwidth]{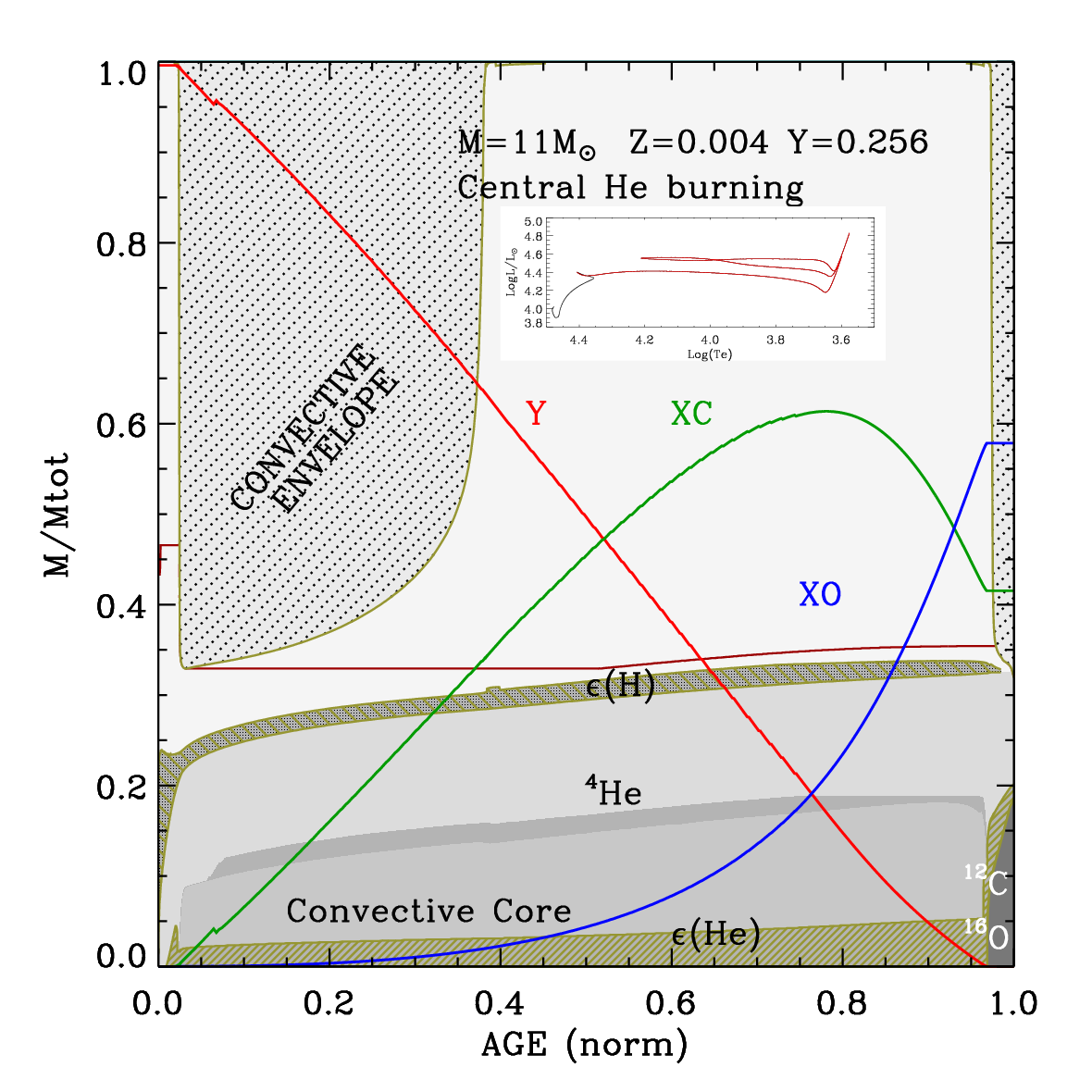}
\caption{Kippenhahn diagrams of central helium-burning model stars
 of $M$=11M$_\odot$ and $Z$=0.004 computed with different envelope overshooting,
EO=0.7\HP, EO=2\HP and EO=4\HP, from left to right respectively. The insets show the corresponding evolutionary tracks in the HR diagram. \label{Psi_laut}}
\end{figure*}
\subsection{Mass-loss rates}
Mass loss is not considered for masses below 12\Msun, in the
{\sl\,PARSEC}~V1.1 tracks, but it cannot be neglected for larger masses.
In this section we describe the algorithm used for the new sets of massive star models,
 obtained by combining
several prescriptions found in literature for the different spectral types.

In the Blue Super Giant (BSG) phases (\Teff $\geq$ 12000~K) we adopt the relations provided by
\citet{Vink_etal00,Vink_etal01}. This formulation, that we name R$_{V01}$, shows an overall dependence of the mass-loss rate on the metallicity,  of the kind
$\dot{M}\propto(Z/Z_\odot)^{0.85}{\rm M_\odot/yr}$.
For the red supergiant stars (\Teff $<$ 12000~{\rm K}) 
 we adopt the relations provided by \citet{de_Jager_etal88}, R$_{dJ}$.
We assume the same dependence on the surface metallicity as in R$_{V01}$.
In the  Wolf-Rayet (WR) phases we use the
\citet{Nugis_etal00} formalism, with its 
own dependence on the stellar metallicity (R$_{NL}$).

A relevant aspect for the more massive stars concerns the transition between O-type 
phase to  the Luminous Blue Variable (LBV) and Red Super Giant (RSG)  phases and finally to the WR phase
and, equally important, the dependence of this transition upon the surface metallicity of the star.
In \citet{Bressan_etal93A&AS_100_647B} as an example of the Padova models,
but generally also in
evolutionary computations made by other groups,
the transition to the super-wind phase
corresponding to the LBV stage
is artificially set when the models cross the \citet{Humphreys_Davidson_1979} instability limit in the
HR diagram.
During this phase, at solar metallicity, the mass-loss rates reach
values of $\dot{M}\simeq$10$^{-3}{\rm M_\odot/yr}$.
In the previous Padova models this limit is set independently
from the metallicity of the galaxy, though the mass-loss rates themselves do depend on the
abundance of heavy elements as indicated above.
More recently however, the behaviour of the mass-loss rates around this transition has been thoroughly
analyzed by \citet{Grafener_etal08} and by \citet{Vink_etal11}.
They find that
the mass-loss rates are strongly enhanced when the stars approach the Eddington limit,
\begin{equation}
   \Gamma=\frac{kL}{4{\pi}cGM}=1
\end{equation}
where the symbols have their usual meaning.
We have thus included in our formalism the explicit dependence of the mass-loss rates from the ratio of the star luminosity
to the corresponding Eddington luminosity, $\Gamma$,
as described by \citet{Vink_etal11} for solar metallicity.
It is worth noticing that, by including this dependence on the mass-loss rates, we are able to reproduce the observed Humphreys-Davidson transition limit at solar metallicity (Chen et al. 2014, in preparation).
However, to extend this behaviour to different galactic environments we need to know how the metallicity affects
mass-loss rate,
near the Eddington limit.
This has been thoroughly analyzed by \citet{Grafener_etal08}
in a broad metallicity range, $10^{-3}~Z_\odot\leq~Z\leq~2~Z_\odot$,
but only for the case of Wolf Rayet stars, and by \citet{muijres_etal12} who investigate the dependence of 
mass-loss rate on the CNO abundances.
The analysis of \citet{Grafener_etal08} shows that
the dependence of the mass-loss rate on the metallicity is also
a strong function of $\Gamma$.
While at low values of $\Gamma$ the mass-loss rate obeys
the relation $\dot{M}\propto(Z/Z_\odot)^{0.85}{\rm M_\odot/yr}$,
at increasing $\Gamma$ the metallicity dependence becomes weaker ,
and it disappears
as $\Gamma$ approaches 1.
In absence of a more comprehensive analysis of the dependence of the mass-loss rates on the surface metallicity and $\Gamma$, we
have adopted the results of \citet{Grafener_etal08}.
In brief, we assume that the mass-loss rates scale
with the metallicity as
\begin{equation}
\dot{M}\propto~(Z/Z_\odot)^\alpha,
\label{mdotgamma}
\end{equation}
with the coefficient $\alpha$
determined from a fit to the published relationships by \citet{Grafener_etal08}
\begin{eqnarray}
   \alpha =  0.85 \qquad \qquad \qquad & & (\Gamma < 2/3) \nonumber \\
   \alpha =  2.45-2.4\times\Gamma \qquad  & & (2/3\leq\Gamma\leq{1})
\label{mdotalpha}
\end{eqnarray}
In summary, our algorithm for the mass-loss rates in different evolutionary phases and metallicities  is the following.
Besides the already specified mass-loss rate formulation
R$_{V01}$ we compute also
the dependence on $\Gamma$, using the tables provided by \citet{Vink_etal11}, R$_{\Gamma}$. However, we scale R$_{\Gamma}$ with the metallicity, using Equation \ref{mdotgamma} and \ref{mdotalpha}.

During the phases BSGs, LBVs and RSGs we adopt the maximum between
R$_{V01}$, R$_{dJ}$ and R$_{\Gamma}$.
When the stars enter
the Wolf-Rayet phase, which we assume to begin when the surface Hydrogen  fraction  falls below $X=0.3$
 and the effective temperature \Teff$>30000~{\rm K}$,
we consider also the \citet{Nugis_etal00} formulation and we adopt the maximum $\dot{M}$
between
R$_{V01}$, R$_{dJ}$, R$_{\Gamma}$ and R$_{NL}$,
after scaling each rate with metallicity using
the appropriate scaling law.
In this way we also obtain a smooth transition
between the mass-loss rates in the different evolutionary phases.


\subsection{The evolution across the HR diagram}
In Figures \ref{fig_HRD_EO_Z001}
and \ref{fig_HRD_EO_Z004} we plot the HR diagram of our new evolutionary tracks of massive stars together with those of intermediate mass stars,
for two different values of the chemical composition, $Z$=0.001 and $Z$=0.004.
In each figure, the three panels from top to bottom, refer to
the three  different
values assumed for the envelope overshooting,
EO=0.7\HP (the standard {\sl\,PARSEC}~V1.1 value), EO=2\HP and EO=4\HP, respectively.
For each evolutionary track,
the central hydrogen burning phase is coloured in blue, the central helium burning phase in green and the contraction phases in red while,
for sake of clarity, the pre main sequence phase is not reproduced.

The black
lines at the highest luminosities show the \citet{Humphreys_Davidson_1979}
limit. This  limit marks the
region  of the HR diagram where, at near solar metallicity,
there are no observed supergiant stars.
The lack of supergiant stars in this region is interpreted as a signature of the
effects of enhanced  mass-loss rates when the star enter this region. This interpretation is supported
by the presence, around this limit, of LBV stars which are known to be characterized by high mass loss rates.
Indeed, the \citet{Humphreys_Davidson_1979} limit is well reproduced by our new models
at near-solar metallicity, due to the
boosting of mass-loss rates near the Eddington limit (Chen et al. 2014, in preparation).
However, at the metallicities investigated in this paper, $Z$=0.001 and $Z$=0.004, the boosting is mitigated by the reduction factor introduced by Eq.~\ref{mdotgamma} and Eq.~\ref{mdotalpha}.
At $Z$=0.001 the upper main sequence widens significantly  and the more massive stars, because of their very large convective cores,
evolve in the ``forbidden''  region even during the H-burning phase. They also ignite and burn central helium as ``red'' super-giant stars.
Notice that the most massive stars begin the main sequence already within the ``forbidden''  region.
At larger metallicity, $Z$=0.004, the widening of the upper main sequence begins at lower luminosity because of the larger opacities
of the external layers but, at the same time, the effects of the mass-loss enhancement near the Eddington limit are more relevant.
For both metallicities, in the most massive stars the mass-loss rates are high enough to peal off  a significant fraction of the
hydrogen rich envelope  and the stars exhaust central Hydrogen near the
\citet{Humphreys_Davidson_1979} limit. Nevertheless, after central He ignition the stars still move
toward the red supergiant phase, until mass loss peals off the entire  envelope and the stars evolve back, toward the Wolf Rayet phase.
This effect is more evident at higher metallicity.

At lower luminosities the intensity of  mass-loss is lower and it cannot more affect the surface evolution of the stars.
There is an intermediate region in the HR diagram where stars ignite helium burning near the main sequence and then evolve toward
the RSG phase. At decreasing initial mass, central He ignition shifts toward the RSG phase and the path of the stars
in the HR diagram becomes similar to that of intermediate mass stars, with the presence of  blue loops during the central helium burning phase.
In general the blue loops are wider for higher  initial mass,
indicating the existence of a bifurcation in the central Helium burning sequences that begins at intermediate-mass stars and
persists up to the massive stars. Together with the main sequence, these sequences are the most
distinct features in the colour magnitude diagrams of star forming galaxies
and constitute the anchors of any study based on the simulated CMD.
On simple grounds they can be explained by considering
 equilibrium structures made by a  helium core and a non negligible outer hydrogen-rich envelope. 
In the HR diagram these
equilibrium structures are expected to lie between the main sequence and the Hayashi
line \citep[][]{Lauterborn_etal71}.
They lie near the main sequence when the helium core is negligible, and near the Hayashi line
when the virial temperature of the helium core, $\propto\Phi=M_C/R_C$ with $M_C$ and $R_C$ being the mass and radius of the helium core,
is larger than a critical threshold.
The latter depends on the mass of the star, the chemical composition, some reaction rates and possibly other input physics
such as internal mixing,
rendering the morphology of the loops quite dependent on several details of stellar evolution
\citep{Iben66,Brunish_etal90,Xu_Li04, Weiss_etal05,Halabi_etal12}.
Such complex dependence makes the theoretical predictions quite difficult, especially at increasing metallicity.
For example at $Z$=0.001 with  canonical envelope overshooting (top panel of Figure \ref{fig_HRD_EO_Z001}), at increasing luminosity
the loops become initially quite extended but then
their extent decreases and thereafter increases again.
This behaviour is even more marked at $Z$=0.004 (upper panel of Figure \ref{fig_HRD_EO_Z004}).
The loops initially become more extended at increasing mass,  but above $M$=6.4\Msun  their extent decreases,
and they  even disappear
for $M$=10M$_\odot$ and $M$=11M$_\odot$.  Above these masses, they fully develop again
until the beginning of central He  burning shifts in the yellow/blue side of the HR diagram.

\subsection{Blue loops: the role of envelope overshooting}\label{hrenvov}
\begin{figure}
\includegraphics[angle=0,width=0.45\textwidth]{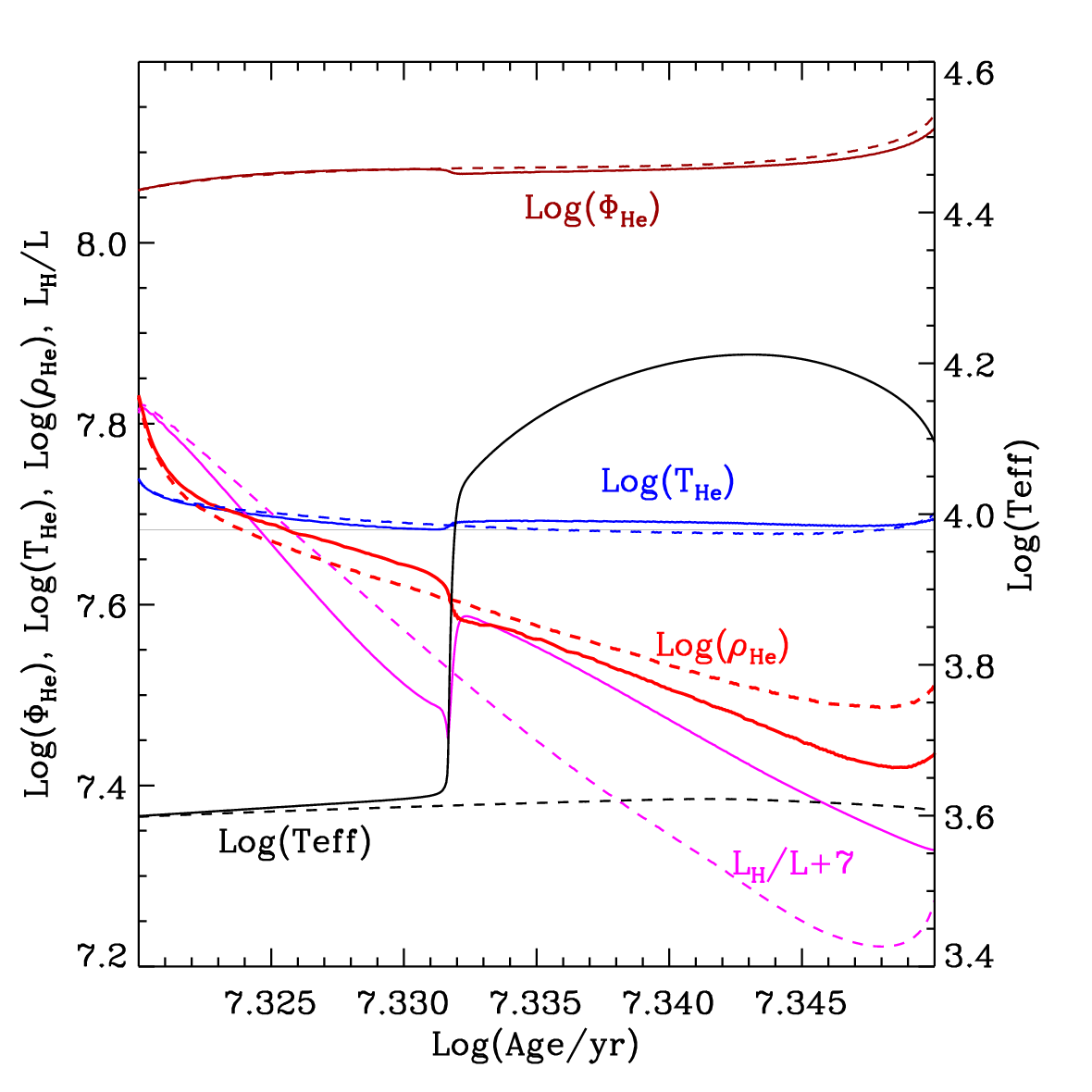}\\
\caption{Evolution of the effective temperature, the H-shell luminosity and other quantities evaluated at the border of the He core, for
the models of Figure \ref{Psi_laut}  with envelope overshooting values EO=0.7\HP (dashed lines) and EO=4\HP (solid lines), respectively. See text for more details.\label{comp_EO07EO4}}
\end{figure}

The reason of the presence or absence of extended blue loops during the central helium burning phase
has been thoroughly investigated in the past.
Among the effects that limit the extent or inhibit the appearance of blue loops,
the most important one is certainly extended mixing from the core during the Hydrogen burning phase,
either due to overshooting \citep{Bertelli_etal85} or to differential rotation
\citep{heger_langer_00, Georgy_etal13}.
This effect can be understood by considering that
it is  more difficult for a star with a larger He core
to decrease $\Phi$ below the critical value needed to begin the loop, because
$\Phi$ increases with the core mass in a virialized structure, for which the radius increases with a power ($\sim$0.5) of the mass.
On the other hand, it is well known that the presence of extended mixing below the bottom of the convective envelope favours the development of an extended loop \citep{Alongi_etal91}. To better clarify this point,
we show in
Figure \ref{Psi_laut} the evolution of the internal structure during the
central helium-burning phase, of a model with
 $M$=11\Msun
and $Z$=0.004, computed with canonical core overshooting and three different values
of the envelope overshooting  EO=0.7\HP ({\sl\,PARSEC}~V1.1), EO=2\HP and EO=4\HP.
The models begin helium burning with the same internal structure,
apart from the larger inward penetration of the envelope convection
due to the different efficiency of the overshooting during the first dredge-up episode, which changes the location of the hydrogen/helium discontinuity (indicated in the figure by the brown horizontal line at $M/M_{\rm{tot}}\sim$0.35). By comparing the three models it turns out that the only noticeable internal difference is
the size of the mass pocket between the H-burning shell and the H-He discontinuity left by the first dredge-up episode.
Only the models with enhanced envelope overshooting (EO=2\HP and EO=4\HP) perform an extended blue loop. The loop  begins when the central hydrogen shell approaches the H-He
discontinuity, in a way that reminds
 us of what happens to a red giant star when its internal hydrogen shell reaches the hydrogen discontinuity.
Because of the high temperature dependence of the CNO reaction rates,
when a discontinuity in the hydrogen profile is encountered, the structure readjusts on
a thermal time-scale toward a slightly lower luminosity and a higher effective temperature
Afterwards the star continues
climbing
 the RGB on the nuclear time-scale,
giving rise to the well known bump in the luminosity function.
What happens in the interiors of the different models of $M$=11\Msun
can be seen in Figure \ref{comp_EO07EO4}, where we compare several quantities
as a function of the evolutionary time during central helium burning.
The dashed lines in the figure refer to the standard {\sl\,PARSEC}~V1.1 model (EO=0.7\HP)
while the solid line to the model with  EO=4\HP (see also Figure \ref{Psi_laut}).
While helium burning proceeds, the temperature and
the density at the border of the He core continuously decrease.
Once the H-burning shell approaches the H discontinuity (brown horizontal line in
Figure \ref{Psi_laut}) in the model with large EO, the density decreases and the temperature increases in a thermal time-scale. In the same time the H luminosity increases from 50\% to
60\% while the star shifts towards the main sequence, in the blue loop.
Notice that  $\Phi$ has a deep during this transition.

With the models of  $M$=10\Msun and $M$=11\Msun
for $Z$=0.004, with canonical core and envelope overshooting ({\sl\,PARSEC}~V1.1),
we also analyze the importance of changing other parameters, such as
the mixing lenght, the $\mathrm{^{12}C(\alpha,\gamma)^{16}O}$ reaction rate
\citep{Iben66,Bertelli_etal85,Brunish_etal90} and
the  $\mathrm{^{14}N(p,\gamma)^{15}O}$ reaction rate \citep{Xu_Li04,Weiss_etal05,Halabi_etal12},
because these parameters are known to affect the extent of the loops.
We find that by modifying these parameters within a suitable  range
the above models are not able to perform extended blue loops in the HR diagram.
We thus conclude that the most critical parameter for the morphology of the He blue loop is the size of the mass pocket between the advancing H-burning shell and the H discontinuity left by the previous first convective dredge-up \citep{Alongi_etal91,Godart13}.
The earlier the shell approaches this discontinuity, the earlier the loop begins and the larger is its extent.
\begin{figure}
\includegraphics[angle=0,width=0.45\textwidth]{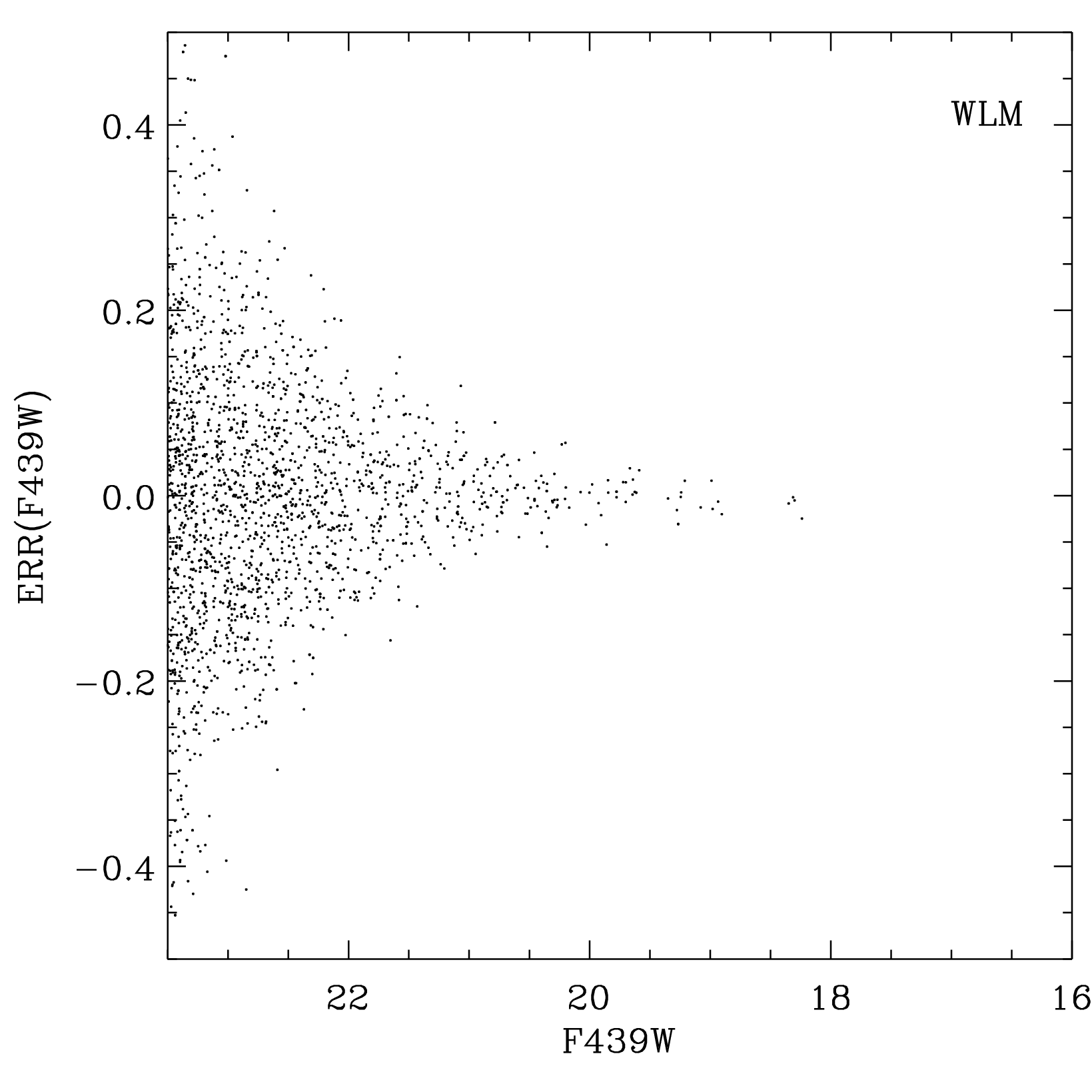}
\caption{Simulated photometric errors as a function of the apparent
magnitude in the F439W filter for WLM stars.
\label{fig_erwlm}}
\end{figure}
\begin{figure}
\includegraphics[angle=90,width=0.45\textwidth]{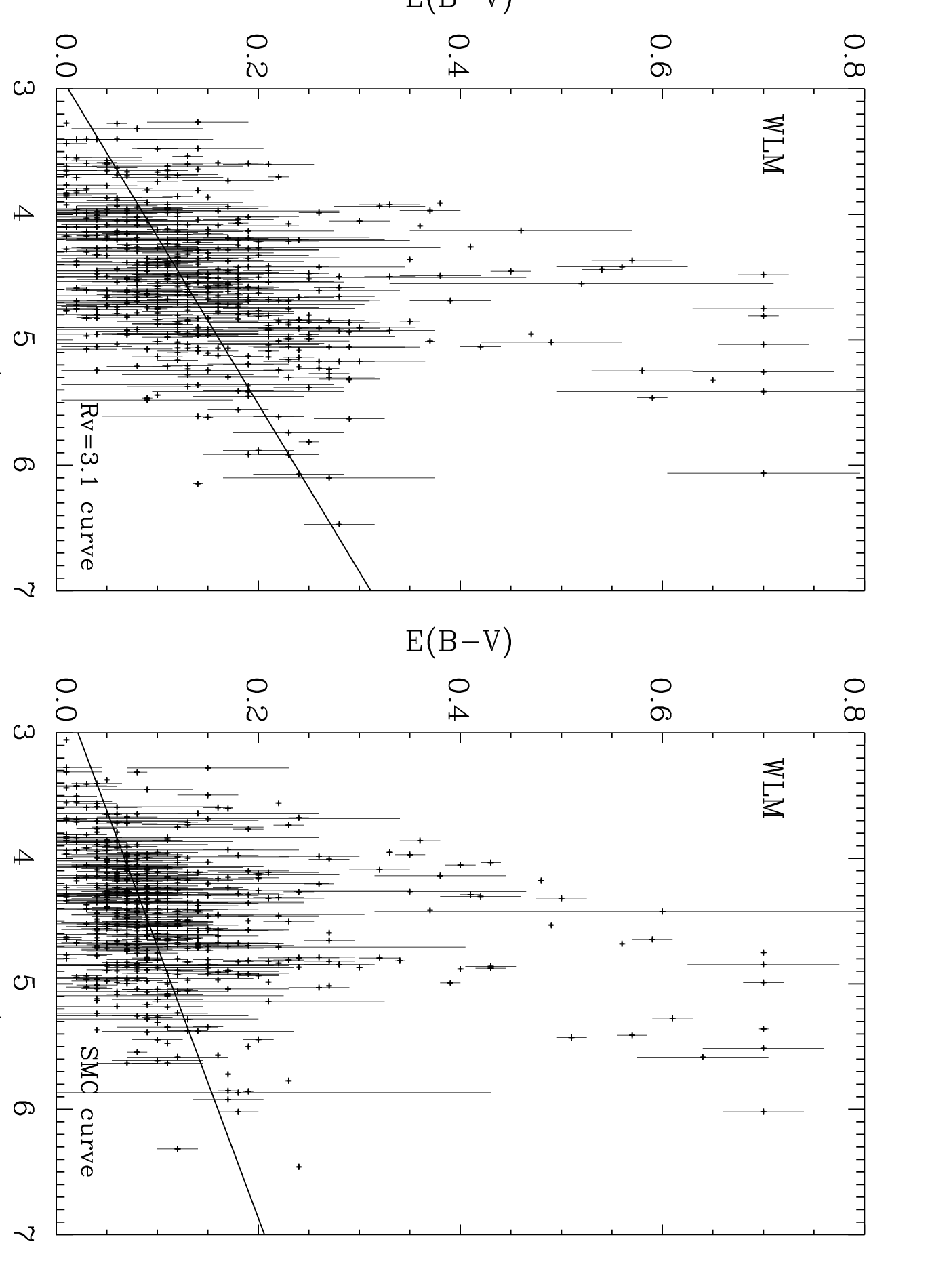}
\caption{Reddening of WLM stars, derived from multi-band SED fitting using the Galactic (left panel)
or the SMC (right panel) extinction curves.
\label{fig_exwlm}}
\end{figure}
For this reason and in view of the comparison of the new models with the observations that will be performed in the next sections,
we  compute additional sets of models with increased amount of
envelope overshooting, i.e. with  EO=2\HP and EO=4\HP.
The stellar evolutionary tracks with increased envelope overshooting are shown
in the middle and lower panels of Figures \ref{fig_HRD_EO_Z001} and \ref{fig_HRD_EO_Z004}
for $Z$=0.001 and $Z$=0.004, respectively.
As expected, after including a larger
mixing at the base of the convective envelope
the blue loops become more extended, though at the lower metallicity
the effect is not large. In this case, only by assuming a 
large mixing scale (EO=4\HP) the effect becomes significant.
It is worth noticing that, though large, values of EO between 2\HP and 4\HP are not uncommon.
For example they are used to enhance the efficiency of the carbon  dredge up during the thermally pulsing Asymptotic Giant Branch phase \citep{Kamath_etal12}.

\section{Data}\label{sec:data}
\subsection{Photometric Data}
To compare our models with observations, we use photometry from  \citet{Bianchi_etal12}'s
Hubble Space Telescope (HST) treasury survey  HST-GO-11079,
which imaged star-forming regions in
six Local Group dwarf galaxies, \object{Phoenix}, \object{Pegasus}, \object{Sextans A},
\object{Sextans B}, \object{WLM} and \object{NGC 6822}, and in M31 and M33.
Multi-band imaging with WFPC2 includes six filters from far-UV to near-IR, F170W, F255W, F336W, F439W, F555W F814W.
\footnote{http://dolomiti.pha.jhu.edu/LocalGroup/}
\citet{Bianchi_etal12} survey was focused on the study of the massive stars in these galaxies, therefore the exposures
were tuned to provide complete filter coverage with high S/N for the hottest stars,
while cooler stars have progressively lower S/N in the UV filters.  We refer the reader to \citet{Bianchi_etal12}
for details on the data reduction and photometric quality;  for the purpose of this work it is sufficient to recall that incompleteness
reaches 20\% at mag=21.0, 22.8, 22.9 and 22.0 in F336W, F439W, F555W, and F814W, respectively.

After careful examination of the CMDs we conclude that the three best galaxies
suitable for  the analysis of their star formation
history to check
 the performance of the new {\sl\,PARSEC}~V1.1 models of massive stars at low metallicity
are \object{WLM},  \object{NGC 6822} and \object{Sextans A}.
The other three galaxies,
\object{Phoenix}, \object{Pegasus} and \object{Sextans B}, are less populated of massive stars and
therefore less suitable for our purpose.
Furthermore, deeper photometry  in the F555W and  F814W filters
 is available for Sextans A
\citep[HST-GO-10915,][]{Dalcanton_etal09}, which we use for
comparison with simulated CMDs.
The CMD of Sextans~A, WLM and NGC 6822 are shown in the upper panels of Figure \ref{SEX_A_SIM},
\ref{WLM_SIM} and \ref{NGC6822_SIM}, respectively.
\begin{figure}
\centering
\includegraphics[width=0.45\textwidth]{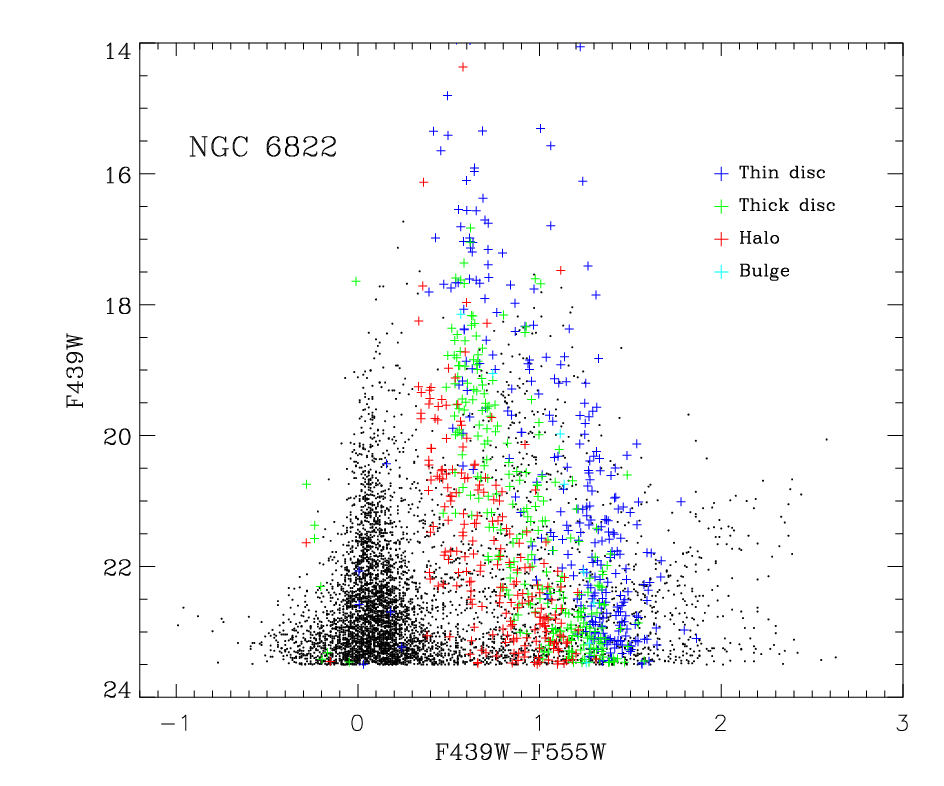}
\caption{Estimated MW contamination in the colour magnitude diagram of NGC 6822. Black dots
 are the observed stars while coloured dots are the predicted
contamination by the four
Galactic components as indicated in the colour code labels.
}\label{fig_contam}
\end{figure}
\subsection{Metallicity of the sample galaxies }
To select  the appropriate metallicity of
the models, we examined the literature for existing measurements of abundances
in our sample galaxies.

\subsubsection{The Metallicity of Sextans A}
An Oxygen nebular abundance of 12 + log(O/H) = 7.49 for \object{Sextans A}, was 
obtained by
\citet{Skillman_etal89}  by means of optical spectrophotometry of $H_{\uppercase\expandafter{\romannumeral2}}$
regions. 
Comparing synthetic and observed CMDs, \citet{Dolohin_etal03_sec}
obtained a mean metallicity of [M/H]=-1.45 $\pm$ 0.2 throughout the measured history of the
galaxy. 
Analysis of three isolated A-type supergiants by \citet{Kaufer_etal04} yielded a 
present-day stellar abundance of the $\alpha$ element Mg as
$\langle[\alpha(Mg_{\uppercase\expandafter{\romannumeral1}})/H]\rangle=-1.09\pm0.02$.
Finally, 
\citet{Kniazev_etal05} presented  spectroscopic observations of three 
$H_{\uppercase\expandafter{\romannumeral2}}$ regions and a planetary nebula (PN) in 
\object{Sextans A}. They derived an average Oxygen abundance of 12 + log(O/H) = 7.54 $\pm$ 0.06 in the
$H_{\uppercase\expandafter{\romannumeral2}}$ regions, which agrees well with \citet{Kaufer_etal04} data.
However, the Oxygen abundance of the PN is significantly higher, 
12 + log(O/H)=8.02 $\pm$ 0.05
and it is explained as self-pollution by the PN progenitor.
Adopting the oxygen solar abundance of 12 + log(O/H) = 8.66 based on the 3D model \citep{Asplund_etal04} and the corresponding total metallicity ($Z$$\sim$0.014),
we obtain for the young populations of  \object{Sextans A} a metallicity of $Z$$\sim$0.001.
For 
this galaxy we will use {\sl\,PARSEC}~V1.1 models with $Z$=0.001.

\subsubsection{The Metallicity of WLM}
For \object{WLM}, \citet{Skillman_etal89_sec} and \citet{Hodge_etal95} measured Oxygen nebular abundances of
 12 + log(O/H) = 7.74 and 7.77, respectively.
\citet{Venn_etal03} found a significant discrepancy between nebular abundance ([O/H] = -0.89) and stellar abundance
[O/H] =-0.21 $\pm$ 0.10. From a study of nine $H_{\uppercase\expandafter{\romannumeral2}}$
regions, \citet{Lee_etal05} obtained a mean nebular oxygen abundance
of 12 + log(O/H) = 7.83 $\pm 0.06$, corresponding to [O/H]=-0.83 when adopting the solar
oxygen abundance of 12 + log(O/H) = 8.66 \citep{Asplund_etal04}.
This is in excellent agreement with the measurement of \citet{Bresolin_etal06} who carried out
a quantitive analysis of three early B supergiants and derived an average 12 + log(O/H) =
7.83 $\pm$ 0.12. Finally, \citet{Urbaneja_etal08} obtained
a weighted mean metallicity [Z/H] = -0.87 $\pm$ 0.06 for several B and A supergiants,
a result 
consistent with the previous measurements.
The metallicity of the young populations is thus [O/H]$\sim$-0.9 which corresponds to
$Z$$\sim$0.0017. Even though 
this value is larger than that of Sextans A,
in the following analysis we will make use of the same models
adopted for Sexstans~A, so that the models with $Z$=0.001 are  compared with  two
bracketing samples of observational data.

\subsubsection{The Metallicity of NGC 6822}
The nebular abundance of \object{NGC 6822} was found to be 12 + log(O/H) = 8.25 $\pm$ 0.07
from a spectroscopic analysis of seven $H_{\uppercase\expandafter{\romannumeral2}}$ regions
\citet{Pagel_etal80}. This result is supported by \citet{Richer_etal95} who
observed 
two planetary nebulae, and deduced 
oxygen abundances 12 + log(O/H) = 8.10 and
8.01, respectively. From a spectral study of A-type supergiants,
\citet{Venn_etal01} derived an average Oxygen abundance of 12 + log(O/H) = 8.36 $\pm$ 0.19 and [Fe/H] = -0.49 $\pm$
0.22 for the  young star population, suggesting also the presence of a gradient in [O/H].
\citet{Tolstoy_etal01} measured the
$Ca_{\uppercase\expandafter{\romannumeral2}}$ triplet in RGB stars and
estimated 
 the mean metallicity of the old stars to be [Fe/H] = -1 $\pm$ 0.5, with values of
individual stars ranging from -2.0 to -0.5. This average value agrees well with \citet{Daviage03} who
obtained the same value from the slope of the RGB. By means of  the ratio between C type and M type
Asymptotic Giant Branch 
stars, \citet{Cioni_etal05} derived a spread of the metallicity
$\Delta$[Fe/H] = 1.56 dex, and \citet{Kang_etal06} found $\Delta$[Fe/H] = 0.07 - 0.09 dex in the bar. Moreover,
\citet{Sibbons_etal12} obtained [Fe/H] = -1.29 $\pm$ 0.07.
The metallicity of  \object{NGC 6822} is definitely higher than that of Sexstans~A and WLM
and therefore
we need to use higher metallicity models for the analysis of its CMD.
In order to favour models with extended blue loops and thus to
avoid any possible problem due to the adoption of a too high metallicity, we will adopt the lowest value compatible with  observations of the young populations. 
Both the lowest values of [O/H] by \citet{Pagel_etal80} and \citet{Venn_etal01}
are compatible with $Z$$\sim$0.0045, again using 12 + log(O/H) = 8.66 and $Z$=0.014 for the Sun.
The simulation of \object{NGC 6822} will thus be performed using  models with $Z$=0.004.

\subsection{Contamination by Foreground Stars}
The contamination of the observed CMDs by
foreground Galactic stars was estimated using the
 TRILEGAL code \citep{2005A&A...436..895G}
via its web-based interface \href{http://stev.oapd.inaf.it/cgi-bin/trilegal}{http://stev.oapd.inaf.it/cgi-bin/trilegal}
The four Galactic components taken into account in TRILEGAL are
the thin Disc, the thick Disc, the halo and the bulge, for which we assume
the default values of the geometrical and constitutive (SFR, metallicity, IMF and extinction) parameters, as specified in the web interface and in the references quoted there.
In particular, the IMF
is assumed to be represented by a
log-normal function \citep{Chabrier01} for all the components.
Extinction is applied to individual stars in the Galaxy,
assuming a dust disk described by a double-exponential decay,
\begin{equation}
A_{V} \propto exp(-|z|/h_{z,dust}) \times exp(-R/h_{R,dust}).
\end{equation}
with default values of scale height $h_{z,dust}=110$ pc,   scale length $h_{R,dust}=100$ kpc 
and a total absorption at infinity  $A_{V}{(\infty)}=0.0378$ mag. In order to calculate the
absorption in each HST filter we use the Galactic extinction curve with $R_{V}=3.1$ \citep{Cardelli_etal89}.
We also take into account the effects of unresolved binary
systems, assuming the default values, i.e.  a fraction of  30\% and  mass ratios uniformly distributed between 0.7 and 1.
We performed TRILEGAL simulations in the direction of the three galaxies used in our analysis,
and scaled the resulting number of stars to the areas covered by the HST survey, given in Table 1 of \citet{Bianchi_etal12}. We extended the TRILEGAL Milky Way model calculations to a limiting magnitude
of 25mag in F439W.
The TRILEGAL simulations indicate
that only in the case of NGC 6822 there is
a significant contamination, as expected
because of its low Galactic latitude.
Figure~\ref{fig_contam} shows the simulated
MW foreground
stars on the CMD of NGC~6822. The contamination has the shape of a plume
around  $m_{F439W}-m_{F555W}\sim{1.4}$ that occupies the region between the blue
and the red sequences of NGC~6822, likely true members of the galaxy.
In order to quantify the contamination, we counted
 the simulated stars
that fall within the box  $0.5 < m_{F439W}-m_{F555W} <1.4$
and $18 < m_{F439W} < 22.5$, in the CMD. In this region
we count 336 stars in the observed CMD while the simulation predicts
 93 stars of the thin disc (blue crosses), 118 of the thick disc(green crosses),
 59 of the halo (red crosses), and 4 of
bulge stars (cyan crosses).
In total we find that there could be 178 possible MW stars in the selected box,
indicating that, in NGC~6822, the MW contamination
above  $m_{F439W}-m_{F555W}>$0.5 is severe and must
be taken into account
in the comparison of the simulated and observed CMDs (Section
\ref{sec:scmd} below).

For Sextans A and WLM the contamination is negligible.

\section{Simulated colour magnitude diagrams}\label{sec:scmd}
To obtain the simulations of the observed CMD,
we modified the code that computes the {\sl\,PARSEC} isochrones,
allowing for the possibility to select a distribution of stars
with suitable star formation history (SFH) and  initial mass function (IMF).

Since our analysis is limited to the  youngest stars,  we do not take into account
an age metallicity relation (AMR) but instead we fix the abundance
to the most likely value of the burst, 
taken from the literature as discussed in Section \ref{sec:data}
The star formation rate is specified as a simple function of time (SFH).
For the IMF,  we assume a simple two power laws representation,
a Kennicutt IMF \citep{Kennicutt_1998}, however with parametrized exponent above 1\Msun.
The SFH and, to a lesser 
degree, the shape of the IMF
are tuned to reproduce the stellar number counts as described below.
To convert luminosity, effective temperature and gravity into magnitude and colours we adopt the bolometric correction tables as in \citet{Marigo_etal08}.
These tables are obtained by convolving large libraries of
stellar spectra with the filters' transmission curves being considered,
as described in \citep{Girardi_etal02}. In the ranges of effective
temperature, surface gravity, and metallicity relevant to this work,
we use the spectra derived from the ATLAS9 model atmospheres \citep{CastelliKurucz03},
which are well calibrated against observations. In the more advanced phases
(Red Supergiants, Wolf Rayet stars) and for the highest masses,
where the effects of mass loss may
be significant, some discrepancies may somewhat affect the transition to
different spectral libraries, but these differences do not alter our
results because these stars are only a minority in the analysed CMDs.

In order
to avoid unnecessary large numbers 
 of simulated stars we define a suitable
absolute limiting magnitude by taking into account the distance modulus of the galaxies
and  the faintest 
magnitude at which the incompleteness reaches  
the $20\%$ in each filter, which is
$\approx$23~mag in F439W and F555W \citep{Bianchi_etal12}.  
For the simulatons of WLM and NGC~6822 we thus select a limiting apparent magnitude of
F439W=23.5 mag. We adopt the same magnitude limit, F555W=23.5~mag,
also for Sextans A, in order to avoid the inclusion of less massive stars in the simulation,
even if the same incompleteness level is reached at fainter magnitudes in the
\citet{Dalcanton_etal09} data.

The outcome of this procedure is a ``clean'' absolute CMD 
that must be translated into the observational
plane adopting the proper distance modulus and
assigning suitable values of error and extinction to each star, as described below. With this procedure we generate,
for each form of SFH and IMF, simulated catalogues
containing 
many more stars than those in the observed CMDs.
The ``simulated CMDs'' are thus drawn from subsets of these large catalogues,
and this procedure is repeated a hundred times to obtain the best fit
luminosity function and its standard deviation, for the selected input parameters.
\begin{figure}
\includegraphics[angle=0,width=0.42\textwidth]{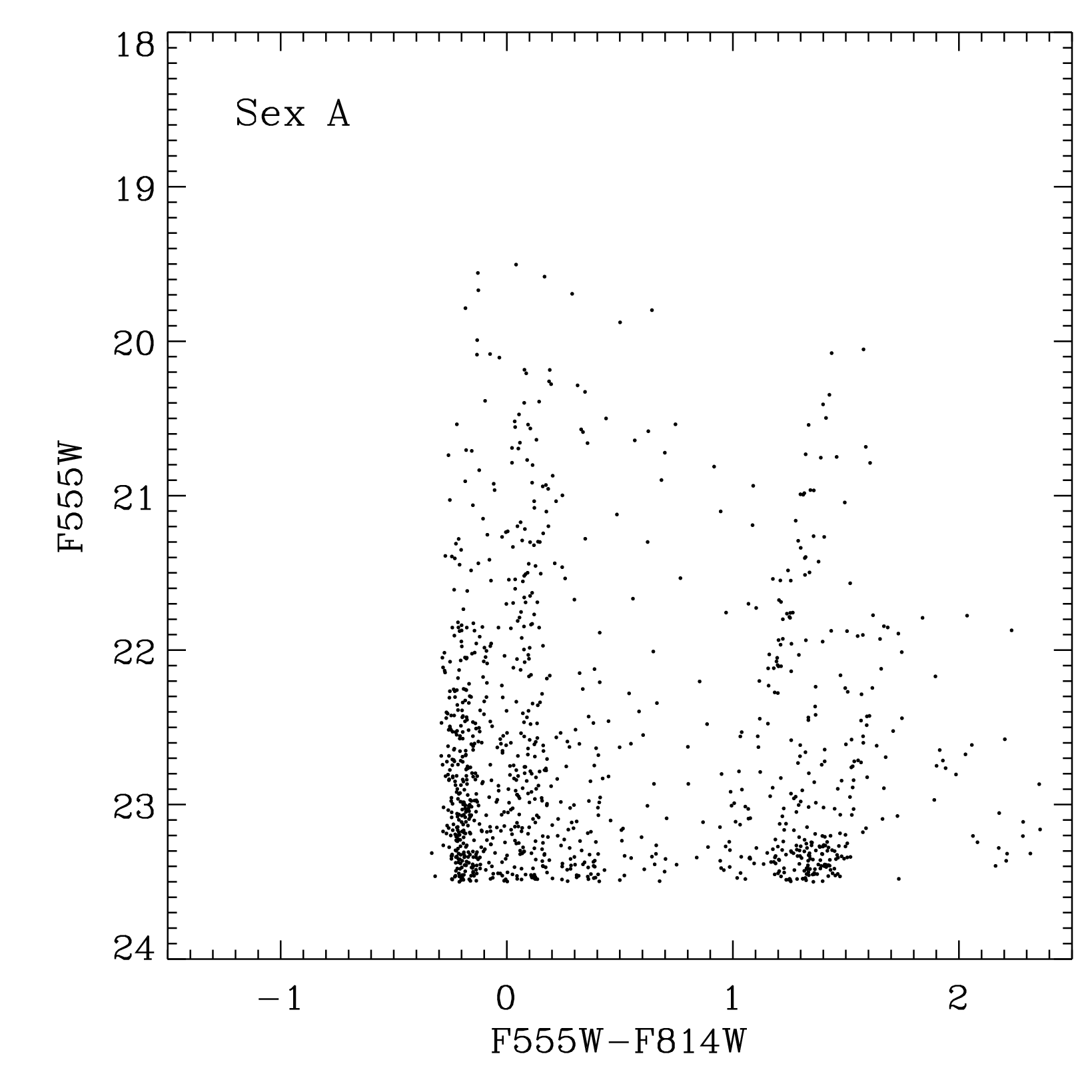}\\
\includegraphics[angle=0,width=0.42\textwidth]{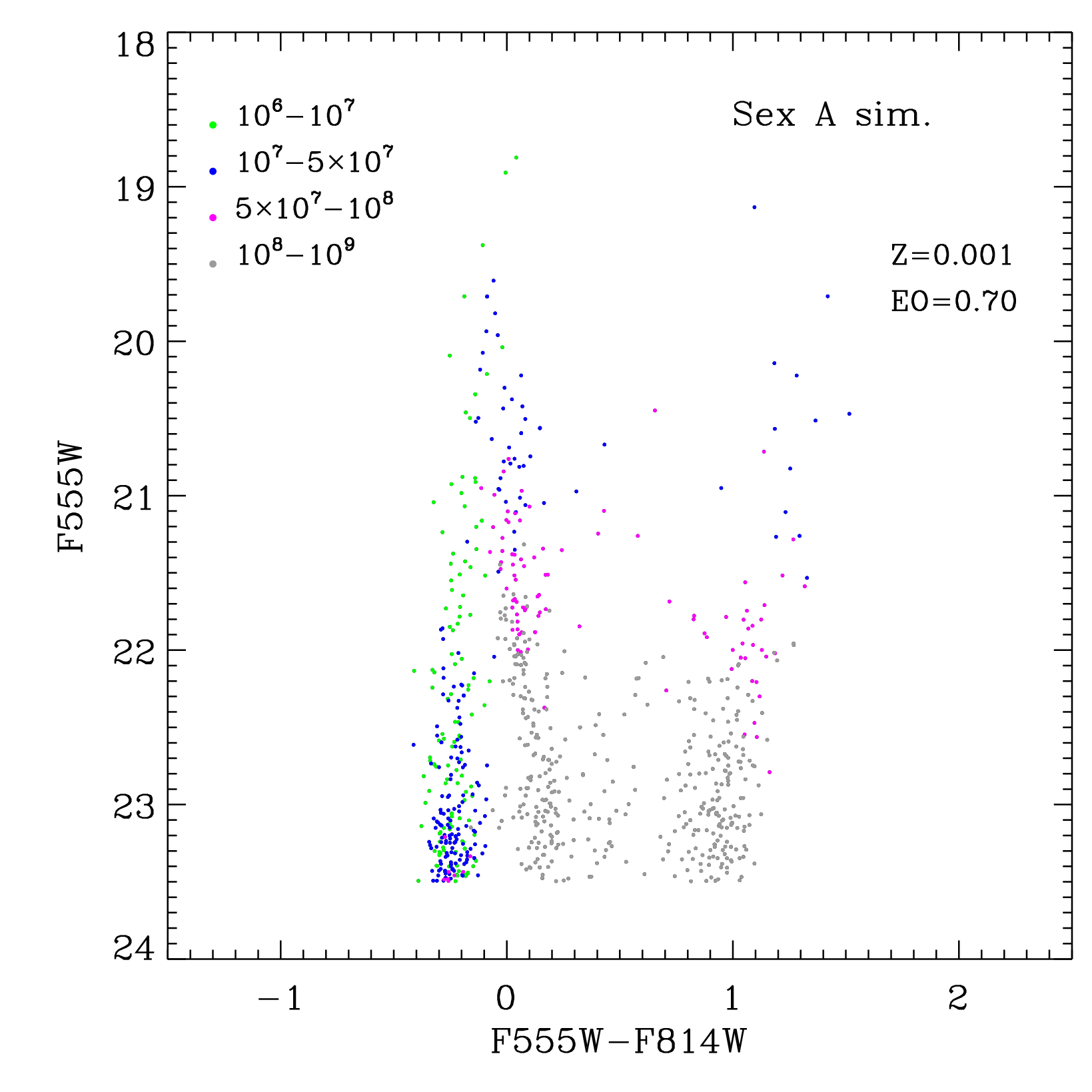}\\
\includegraphics[angle=0,width=0.42\textwidth]{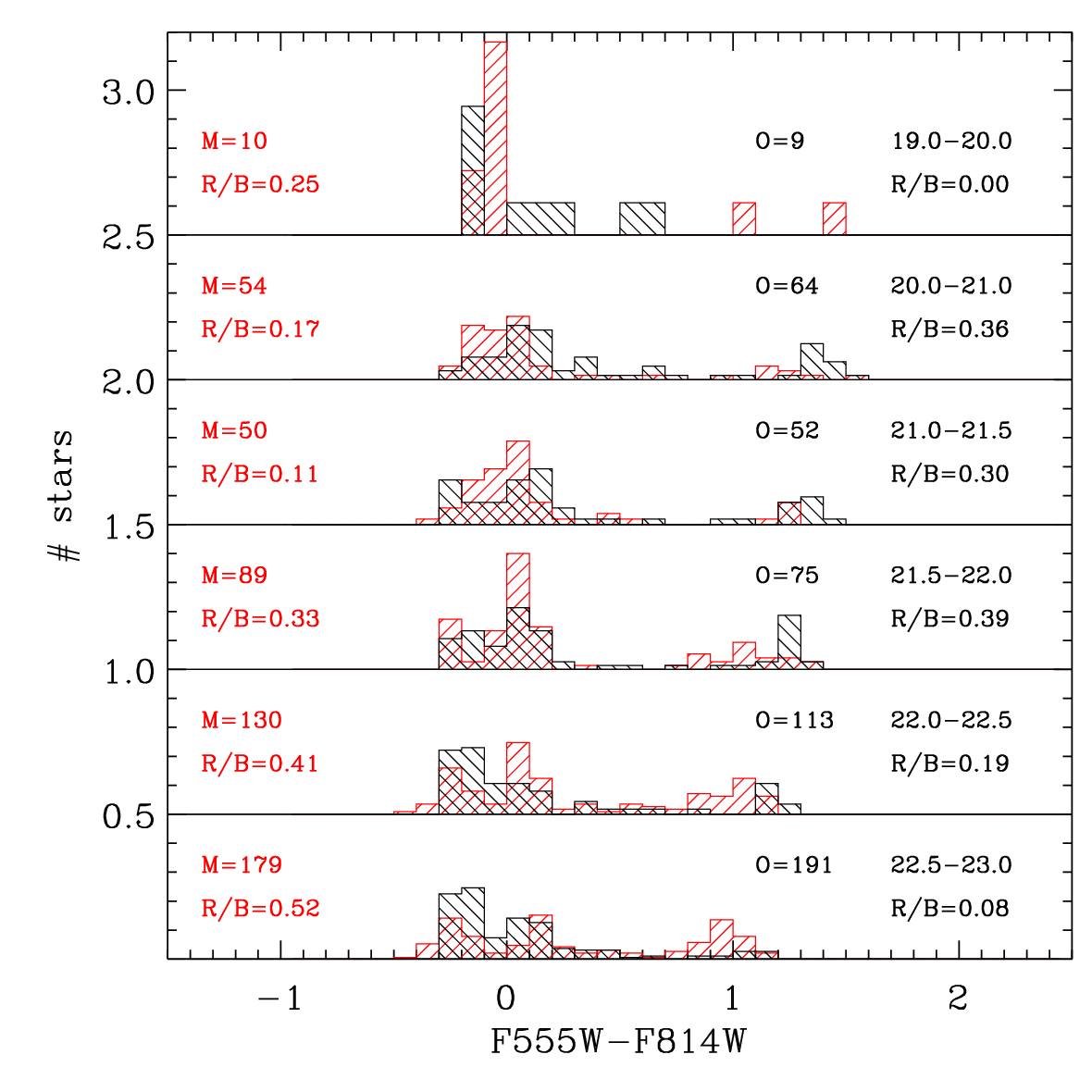}
\caption{Upper panel: Observed color-magnitude diagram of Sextans A. Middle panel: the best model obtained with {\sl\,PARSEC}~V1.1.
Lower panel: comparison of the observed (black)
and modelled (red)  colour distributions.
\label{SEX_A_SIM}}
\end{figure}
\subsection{Simulated errors}
The simulated photometric errors have been derived from the distribution of the observational errors as a function of the
apparent magnitude \citep{Bianchi_etal12, Dalcanton_etal09}, as follows.
For WLM and NGC~6822 we select sources
with HSTphot $type = 1$ and $\mid sharpness\mid < 0.3$ and we plot error {\it vs} 
 apparent magnitude diagrams. These figures are similar to those shown in \citet{Bianchi_etal12} and are not reproduced here. To estimate individual errors for the simulated stars in a given filter, we re-bin the error distribution in 0.1 magnitude bins
and, within each magnitude bin, we evaluate the median error.
Then, the error assigned to each model star
of a given apparent magnitude is randomly drawn from a Gaussian distribution with a
standard deviation derived from the median value corresponding to its magnitude.
An example of the simulated error in the F439W filter for WLM is shown in Figure \ref{fig_erwlm}.
For Sextant~A the errors are determined in the same way,
but using the data from \citet{Dalcanton_etal09}.

\subsection{Accounting for extinction}\label{sec:extinction}
In the analysis of the CMDs of stellar systems the extinction is generally
treated as a constant free parameter that must be derived from a 
fitting procedure.
In this respect, a key feature of this investigation is that we may exploit the information carried by the multi-band photometric coverage of the observations
because our galaxies were observed in six
HST bands by \citet{Bianchi_etal12}. 
For many of the observed stars the
photometry spectral energy distributions (SED)
can be modelled to obtain
extinction, effective temperature and bolometric luminosity at once.
We use such results to define
the attenuation on a star by star basis, in a statistical way.

We use  extinction results from
\citet{Bianchi_etal12}'s SED fitting of the multi-band HST photometry
with theoretical spectral libraries of different metallicities,
$Z$=0.0002, $Z$=0.002 and $Z$=0.02 and
different extinction curves,
``average'' MW  with $R_v=3.1$ \citep{Cardelli_etal89}, ``LMC2'' \citep{Misselt_etal99} or ``SMC'' \citep{Misselt_etal98}.
We restricted this analysis to sources with photometric accuracy better than 
0.25, 0.20, 0.25, 0.10, 0.10,
 in the filters F170W, F225W, F336W, F439W, F555W 
respectively, and applied no error cut in F814W. Such sample restriction to stars detected in
at least five filters,  with small photometric errors in the optical bands, ensures that enough measurements are
available for each star for deriving concurrently two free parameters (\Teff and E(B-V) ),
and  biases
the analysis sample towards the hottest, most luminous stars, because we require detection even in
the far-UV filter.  Results from this survey and similar ones showed  that the hottest stars are usually found in regions of high IS extinction \citep[see e.g.][Figure 12]{Bianchi_etal12}, as we shall discuss again later.
The resulting extinction depends significantly on the adopted attenuation curve
 \citep[see also][]{Bianchi_etal12} but
much less on the adopted metallicity of the theoretical spectra.

As an example, we show in Figure \ref{fig_exwlm} the extinction derived for the stars in WLM by \citet{Bianchi_etal12}.
The two panels show two different cases obtained with the same spectral library, $Z$=0.002,
and two different extinction laws, the Galactic (R$_V$=3.1) one in the left panel and the SMC one, in the right panel, respectively. For each star,
the E(B-V) is plotted against the stellar bolometric luminosity,
which is also derived from the best fit of the observed SED, assuming the distance of the galaxy.
The vertical error bars depict the uncertainties, derived in the SED-fitting process.
In addition to the formal uncertainties, there may be a slight bias due to poor calibration of the
WFPC2 UV filters (CTE corrections) as discussed by \citet{Bianchi_etal12_sec}.
Figure \ref{fig_exwlm} shows 
a trend of increasing E(B-V) at increasing luminosity.
The line shown in the figure is a simple best fit to the data obtained after
selecting only those  with a relative error $\leq$0.2 and
 excluding the outliers (using LADFIT IDL procedure).
The lines shown in the panels show a clear positive slope with
the stellar intrinsic luminosity. The slope obtained with the Galactic extinction law (left panel) is about 50\%
larger than that obtained with the SMC extinction law.
We also observe a significant dispersion at all luminosities and, in general, the more luminous stars have smaller uncertainties.
Of course the derived extinction includes the contribution arising from the MW
which should be almost constant in a given field though we notice that,
at the lower luminosities, there are values that are
even less than the contribution of the Galaxy.
However,
within the
uncertainties, they are compatible with the foreground value, which is significant only
for NGC~6822.
Instead we suppose that the dependence of the additional attenuation with the intrinsic luminosity is a clear signature the internal extinction being
age-selective \citep{Silva_etal98}. In this case the trend arises because younger stars are embedded in a more opaque interstellar medium,
independently from their mass and luminosity
while, older stars, which are generally less massive and of lower luminosity may be less attenuated.
Another noticeable  feature is that in the higher luminosity bins the errors are small and the observed values
indicate that there can be significant star to star variation in the extinction.
There are also  a few stars with small error bars and
significantly higher than average extinction.  These are generally stars
cooler than the main sequence ones
and a likely explanation for their large extinction
is that they could be affected by dust produced by their own circumstellar envelopes. Since there are only a small number of such stars,
they are excluded from the fits shown in the above panels, but this point could deserve further investigation
because it could be an evidence of pristine dust production in low metallicity galaxies.
In summary,
 there is evidence that in our galaxies the extinction
may grow with intrinsic luminosity  for the hot stars,
and may have a significant
star to star variation.
In order to account for these effects in the simulated CMD, we
randomly distribute the extinction values as a function of the (known) intrinsic luminosity of our model stars, by reproducing the best fit relation and the observed dispersion with a Gaussian model.
This effect introduces a further dispersion in the simulated CMDs,
that adds to that of the photometric errors and possibly a
tilt in the modelled stellar sequences. Because we modeled the extinction distribution on
a sample of hot stars, which are mostly associated with regions of high extinction, we should keep in mind in the following discussion that, if cooler stars outside
 the star-forming regions are included in the catalog, the model extinction distribution
may be over-estimated for them;
this is not the case for the \citet{Bianchi_etal12} survey which mostly targeted conspicuous star-forming regions.

For  Sextans~A we
fit the star to star attenuation model derived from
the multi-band
data by \citet{Bianchi_etal12}, and apply it to the
F555W, F814W deeper
data by \citet{Dalcanton_etal09}, using the selected extinction law. Since in the considered range of magnitudes the errors
are negligible in the \citet{Dalcanton_etal09} photometry,
the variable extinction is practically
the only source of star to star dispersion
in the simulated CMD of Sextans~A.

\section{Results with canonical models}\label{sec:results}
As already anticipated in section \ref{sec:tracks}, the simulations of the CMDs of the three galaxies are based on stellar evolution models that use different values of the envelope overshooting.
We begin by discussing the results obtained by using the standard value adopted in
{\sl\,PARSEC}~V1.1, i.e. an envelope overshooting of EO=0.7\HP.
\begin{figure}
\includegraphics[angle=0,width=0.42\textwidth]{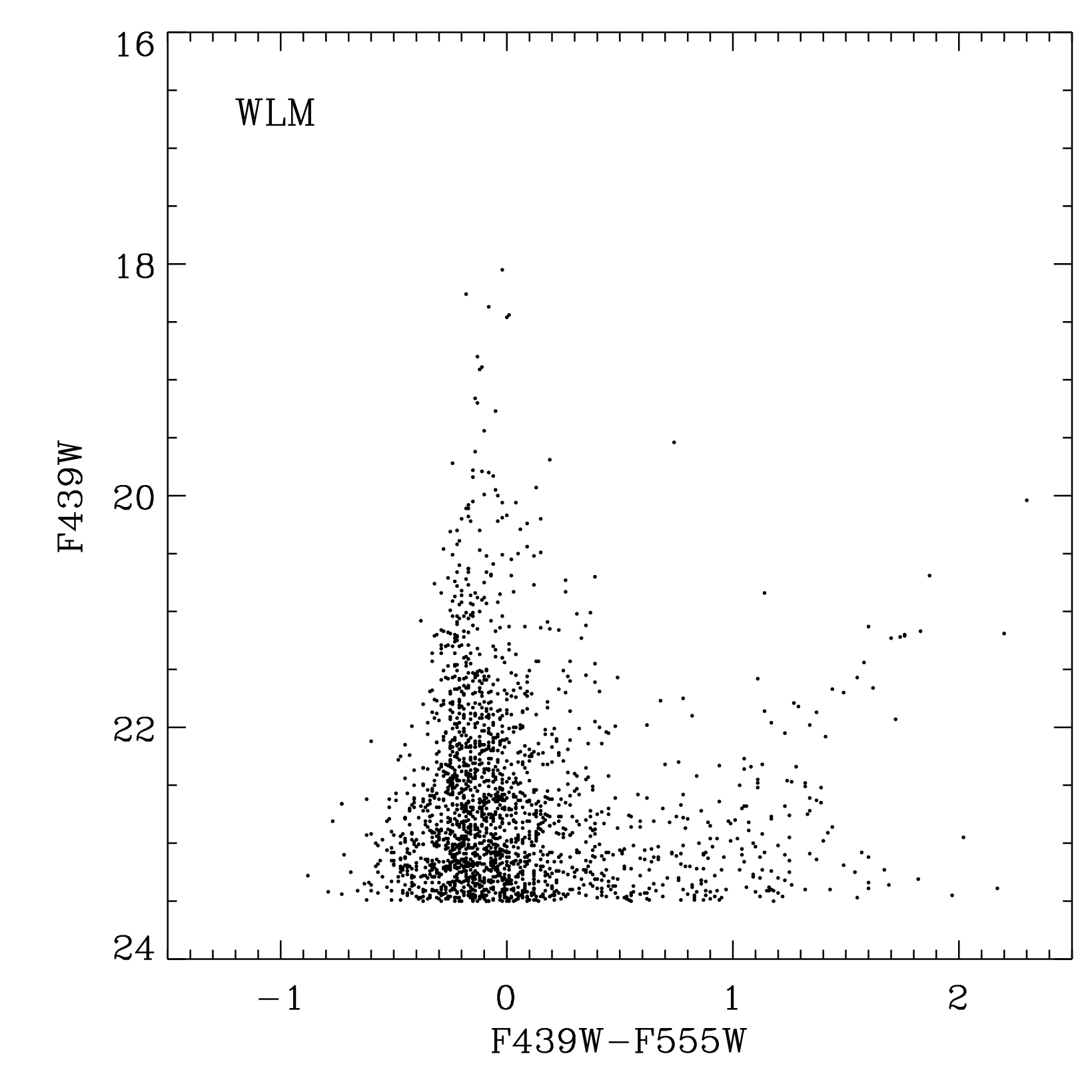}\\
\includegraphics[angle=0,width=0.42\textwidth]{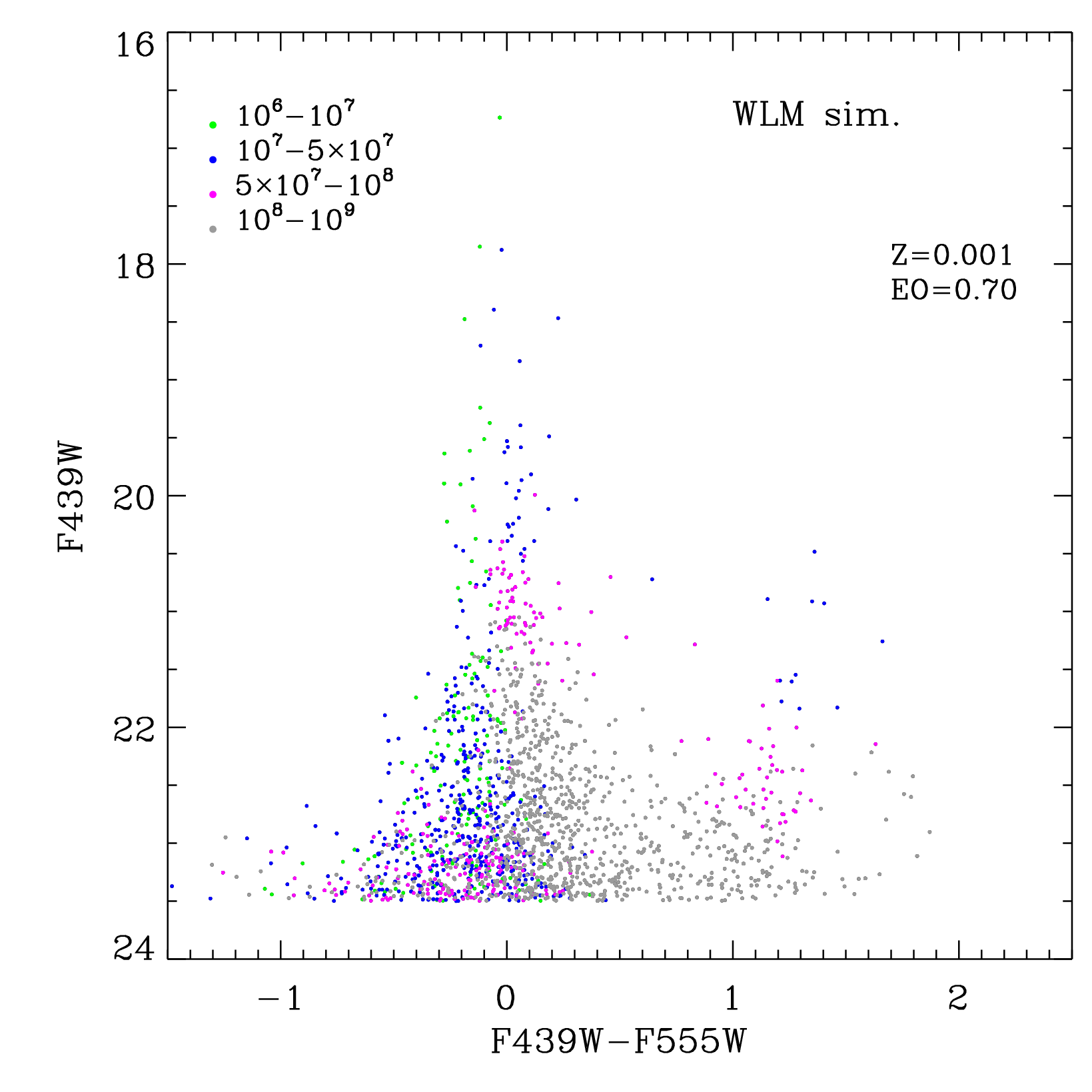}\\
\includegraphics[angle=0,width=0.42\textwidth]{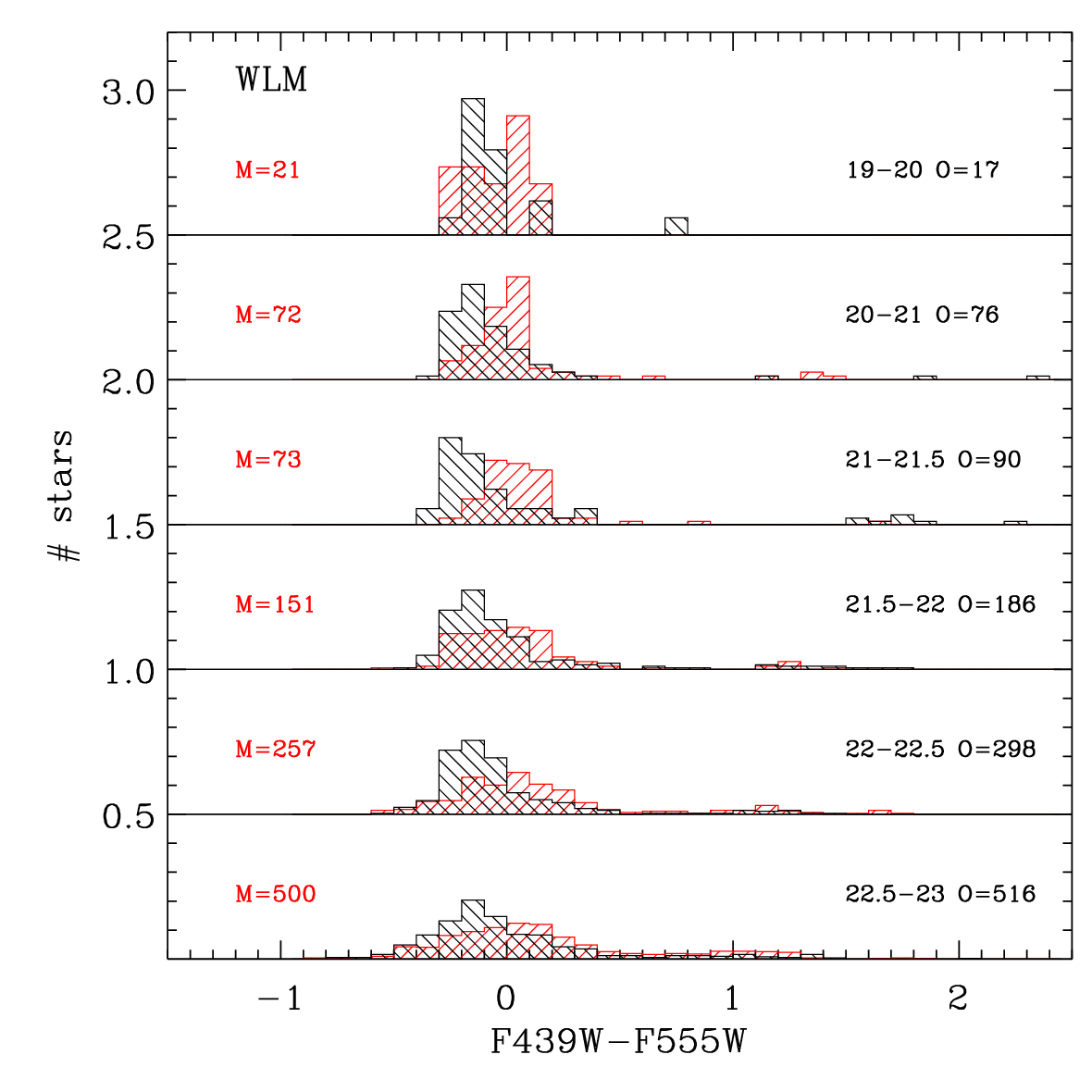}
\caption{Upper panel: Observed color-magnitude diagram of WLM. Middle panel: the best model obtained with {\sl\,PARSEC}~V1.1.  Model stars are colour-coded by age as indicated in the legend.
Lower panel: comparison of the observed and modelled colour distributions. \label{WLM_SIM}}
\end{figure}
\subsection{Sextans A}

\subsubsection{The colour magnitude diagram}
The observed CMD of Sextans A, reproduced in the upper panel of  Figure \ref{SEX_A_SIM},
shows the following noticeable features.
There is a well defined main sequence with stars brighter than F555W=23.5 (our selected magnitude limit)
and bluer than F555W-F814W $\leq$-0.1.
The sequence appears to  become broader and redder at brighter magnitudes.
This effect is unusual because, in general
the sequence
 becomes bluer and narrower at brighter magnitudes, due to the smaller photometric errors.
A parallel sequence, redder by about 0.2-0.3 mags, is also visible and likely corresponds
to the stars that are burning helium in the blue loops. This may not be the case for the brightest stars
that, as shown in the HR diagrams may ignite central helium before reaching the red supergiant phase.
A red sequence, starting at F555W-F814W $\sim$1,  indicates the presence of red giant/supergiant stars
in the red phase before the loop.
Only a scarce number of stars populate the region
between  the red and blue helium-burning sequences,  as expected because of the 
short  crossing timescales of the Hertzsprung gap.
The stars found in this gap are likely yellow supergiants, since the MW contamination that mainly affects this region of the CMD is low for this galaxy.
The RGB and AGB phases of low and intermediate mass make up the sequences at even redder colors (F555W-F814W $\geq$1.2).

For the simulations of Sextans~A we adopt a distance modulus (m-M)$_0=25.61$ \citep{Dolphin_etal03}.
The best simulation obtained with {\sl\,PARSEC}~V1.1 models is shown in the middle panel of 
Figure \ref{SEX_A_SIM}.
This simulation has been selected from a hundred stochastic
realizations made with the same parameters, by means of a merit function based on the differences between
the observed and simulated luminosity function.
Stars of different ages are shown with different colors in the simulated CMD,
as indicated in the corresponding labels. The metallicity adopted for the simulation is $Z$=0.001.
The simulated CMD reproduces  qualitatively all the main features of the observed diagram, except for the
stars redder than F555W-F814W $\geq$1.2 and the sequences that depart from there.
This is because we impose a cut in the simulation
at masses below 1.9\Msun, since we are interested in the recent star formation history
of the galaxy.
For the same reason we discarded from the  analysis below all stars in the observed diagram redder than
the line F555W-F814W = (31.2-F555W )/7 which should correspond to the old population.

\subsubsection{The Star Formation Rate}
Since our simulated CMD catalogue is constructed by imposing a total number
of living stars brighter than a given threshold luminosity, a few steps are needed to derive the normalization of
the SFR  that corresponds to the parameters given in
Table \ref{tab_SIM_SEXA}. In particular we have to take into account
that our  model
catalogue is a super-set of the observed CMD, to which it should be matched. 
So,  we first need to obtain the SFR normalization
of the parent catalogue and then
scale this normalization to the number of stars in the observed CMD. 
For the
first step we selected a small mass interval for which
the stellar evolutionary times are larger than the oldest age of our simulation.
Since all such stars are still alive, they must all be in the catalogue
and they can be easily counted. On the other hand
their number must correspond to the double integral, over the full time interval
and the selected
mass interval, of the product of the SFR and the IMF.
From the comparison between the counted stars and the analytical result
of the double integral, we derive the normalization of the
model SFR used to generate the large catalogue.
To obtain the real SFR from the observed CMD,
we simply scale the model SFR
by imposing that the model
contains the same number of stars as
the observed CMD,
in the magnitude range 
 between 23 and 21.5.
We use this magnitude interval to limit the
stochastic effects introduced by the brightest stars.
Nevertheless, since the normalization
is made after the simulated magnitudes have been assigned
attenuation and
errors, the star formation rate so derived may  slightly change between different models,
even if the underlying parameters are the same.
It is worth noting
that the normalization obtained in this way
may also be affected by all those effects that may modify
the luminosity function in a given pass-band.
One such effect, as we will see, is the variation of the lifetime
of stars in the blue and red sides of the loops.
This effect may change the number of stars in the magnitude bins and so
the resulting
 luminosity function.
Another possible effect is that of the unresolved binary stars
(even if not physical), which
artificially increase
the number of stars
in the
brighter magnitude bins. 
In this respect, HST/WFPC2's resolution (0.1$''$ spatial scale, i.e. less than half pc in WLM and Sextans~A, and about one fourth of a pc projected on the sky, at the distance of NGC~6822)
offers a critical advantage with respect to ground-based observations with typical seeing
of $\gtrsim$1$''$.
Quantification of the gain afforded by HST imaging in resolving individual stars in these
galaxies can be seen, for example, in Figure 3 of \citet{Bianchi_etal12}.  This is particularly relevant for studies of hot stars, which are found in crowded regions because of their young age, the OB associations having not yet
dissolved. The very bright stars are rare, but at B and V magnitudes fainter than $\approx$20, the effect
of unresolved stars in ground-based catalogs, and the gain from HST, become very significant.
We finally note that for random superposition (along the line of sight) the combined photometry of two likely different types of stars will result in an SED hard to fit with a single star, and the source may be discarded from the analysis with quality cuts.

We find that the SFR increases toward recent times ($\tau<$0, Table \ref{tab_SIM_SEXA})
with a characteristic time scale $\mid\tau\mid$=2E8${\rm yr}$.
The average star formation rate in the last 100 million years is
2.5$\times$10$^{-3}{\rm M_\odot/yr}$, as indicated in Table \ref{tab_SIM_SEXA}.
The maximum mass obtained from the simulation is $M\sim$80\Msun.
However, with F555W$\sim$20.5~mag, this is not the brightest star
which instead is an evolved star
with $M\sim$30\Msun and F555W$\sim$18.8~mag.

\begin{table}
\begin{center}
\caption{Parameters of the CMD simulations of Sextans A \label{tab_SIM_SEXA}}
\begin{tabular}{llllll}
\hline\hline
EO & T$_i[{\rm yr}]$ &T$_f[{\rm yr}]$ & $\tau[{\rm yr}]$ & X$_{IMF}$ & $<$SFR$>[{\rm M_\odot/yr}]$ \\
\hline
0.7 &  1E6 &1E9 &-2E8 & 2.35 &  2.5E-3\\
2   &  1E6 &1E9 &-2E8 & 2.35 &  2.9E-3\\
4   &  1E6 &1E9 &-1E8 & 2.35 &  3.2E-3\\
\hline
\end{tabular}
\end{center}
\end{table}
\begin{table}
\begin{center}
\caption{Parameters of the CMD simulations of WLM\label{tab_SIM_WLM}}
\begin{tabular}{llllll}
\hline\hline
EO & T$_i[{\rm yr}]$ &T$_f[{\rm yr}]$ & $\tau[{\rm yr}]$ & X$_{IMF}$ & $<$SFR$>[{\rm M_\odot/yr}]$ \\
\hline
0.7 &  1E6 &1E9 &-5E8 & 2.65 &  3.7E-3\\
2   &  1E6 &1E9 &-5E8 & 2.35 &  2.7E-3\\
4   &  1E6 &1E9 &-2E8 & 2.35 &  2.9E-3\\
\hline
\end{tabular}
\end{center}
\end{table}
\subsubsection{The colour distribution}
In order to make the comparison between the observed and simulated CMDs
more quantitative, we plot in the lower panel of Figure \ref{SEX_A_SIM}
the colour distribution of the stars in different bins
of F555W 
magnitude, 19-20,20-21,21-21.5,21.5-22,22-22.5,22.5-23.
The black histograms refer to the observed CMD while the red histograms are for the
simulated CMD. On the right side of each histogram we show the corresponding magnitude interval,
the number of stars observed  in the interval (O) and the observed ratio (R/B)  of stars redder and bluer than a given threshold colour.
On the left side of the histogram we show the same quantities, number of stars and number ratios (M and R/B), as predicted by the simulations.
For Sextans A we adopted a threshold value of F555W-F814W $=$0.6 and, as already specified,
we did not considered stars redder than the line F555W-F814W = (31.2-F555W )/7, that more likely
belong  to older populations.

The colours and in particular the width of the simulated main sequence
match fairly well those of the observed one, in all the magnitude bins.
Two points are worth discussing.
First we note that, since we are using a
single metallicity,  the observed width of the main sequence
should be ascribed mainly to photometric errors, because the
intrinsic width at 22 mag is $\delta(F555W-F814W)_0\lesssim$0.07~mag.
But the photometric errors
of the observations are very small 
at these magnitudes.
Second, the simulated main sequence is almost vertical in the diagram while,
as already said we would 
expect this sequence to be inclined toward
bluer colours at brighter magnitudes.
Both effects are due to the extinction.
The extinction was evaluated for individual stars 
from simultaneous fits for [\Teff, E(B-V)] of  multiband photometry \citep{Bianchi_etal12}
by means of suitably attenuated spectral models.
As 
discussed in section \ref{sec:extinction}, \citet{Bianchi_etal12}
find that E(B-V), besides  showing a significant dispersion,
is higher for the hot, young stars. 
The dispersion in the attenuation is intrinsic and  due to region to region
variation. Instead its rise with luminosity is  likely caused by the following 
 bias.
All stars are born in region of relatively higher attenuation and
as the time elapses, the parent clouds dissolve and the attenuation decreases \citep{Silva_etal98}.
It is well known that the emission lines (signature of
 young massive stars) in star-bursts
show an attenuation which is about twice that of the optical continuum (mostly from older, lower mass stars) \citep{Calzetti_etal94}.
For a continuous star formation it is thus expected that hot
massive stars (that have been
recently formed)
are the more attenuated, while stars of lower luminosity
(whose ages can span the full star-burst period)
can have any attenuation. Thus age dependent attenuation provides at once the dispersion and
the trend with luminosity that are at the basis of the observed spread and tilt
of the main sequence.

The observed blue helium-burning sequence is also fairly well reproduced 
by the simulation.
However we notice  that
significant discrepancies remain in the colour distribution.
In the three most populated bins, at magnitudes fainter than F555W$\sim$21.5,
the model distribution is clearly
bimodal.
All the simulated diagrams show a gap between the main sequence and the
blue helium-burning sequence that instead is much less evident 
(if any) in the observed CMD.
This gap cannot be filled by invoking errors or 
by varying the attenuation. 
The presence of the gap is an indication the
models are not able to produce extended loops during the helium burning phase.

The sequence of red stars is also fairly well reproduced.
We notice that the slope of this red sequence is
fairly well reproduced because of the trend of the attenuation to be larger
at larger luminosities, i.e. its slope is partially due to
varying  reddening across the sample. 

As far as the number distribution between red and blue stars
(R/B ratio) is
 concerned, we notice significant discrepancies between the
simulated and the observed ratios.
Apart from the highest bin where no red stars are observed,
the R/B ratio is underestimated by the model
below F555W=22, while it is largely overestimated at fainter magnitudes.

In summary, while the morphology of the simulated CMD is very similar to that of the observed
one, inspection of the colour distribution shows that there are some important discrepancies,
mainly concerning
the evolution during the central helium burning phase.

\subsection{WLM}
\begin{figure}
\includegraphics[angle=0,width=0.42\textwidth]{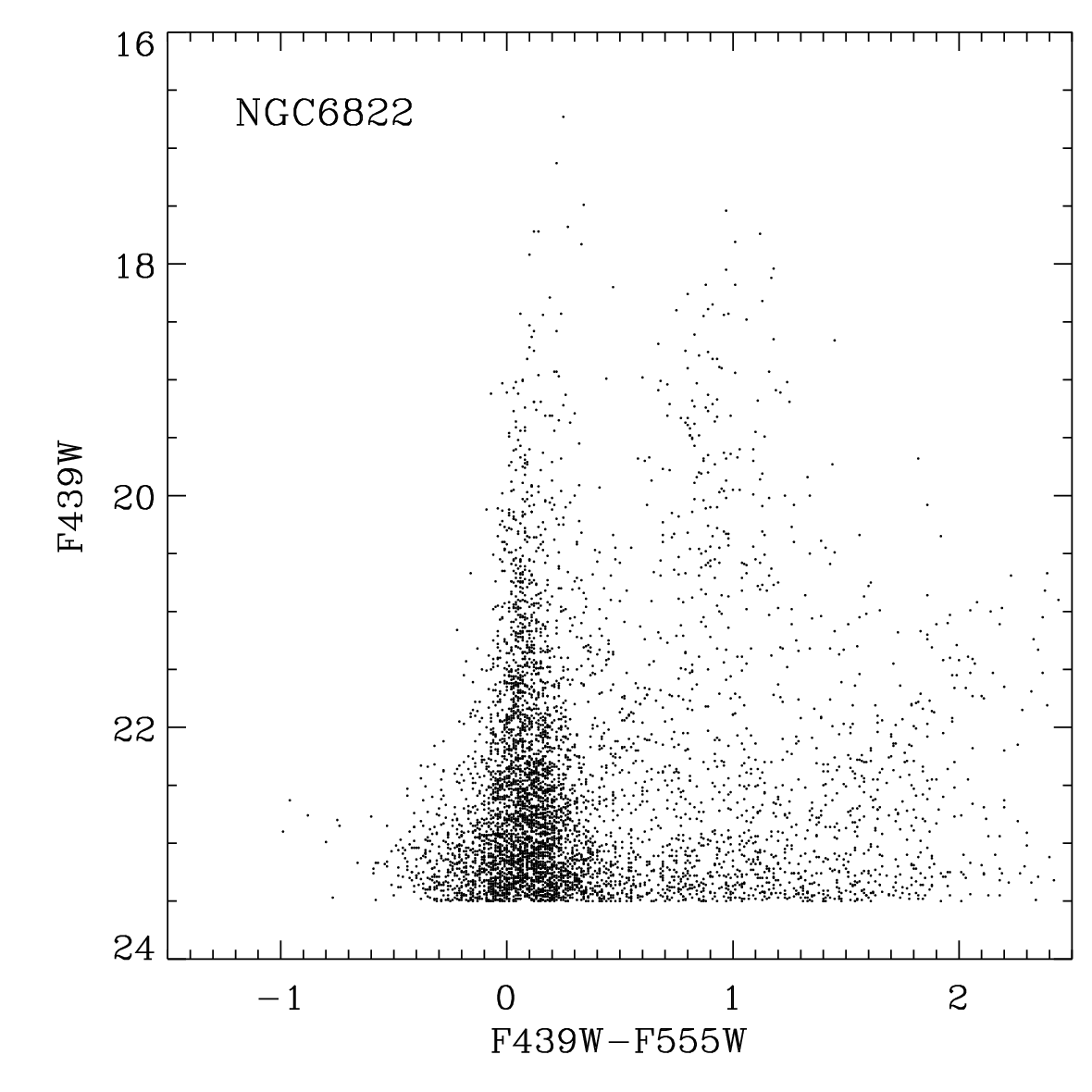}\\
\includegraphics[angle=0,width=0.42\textwidth]{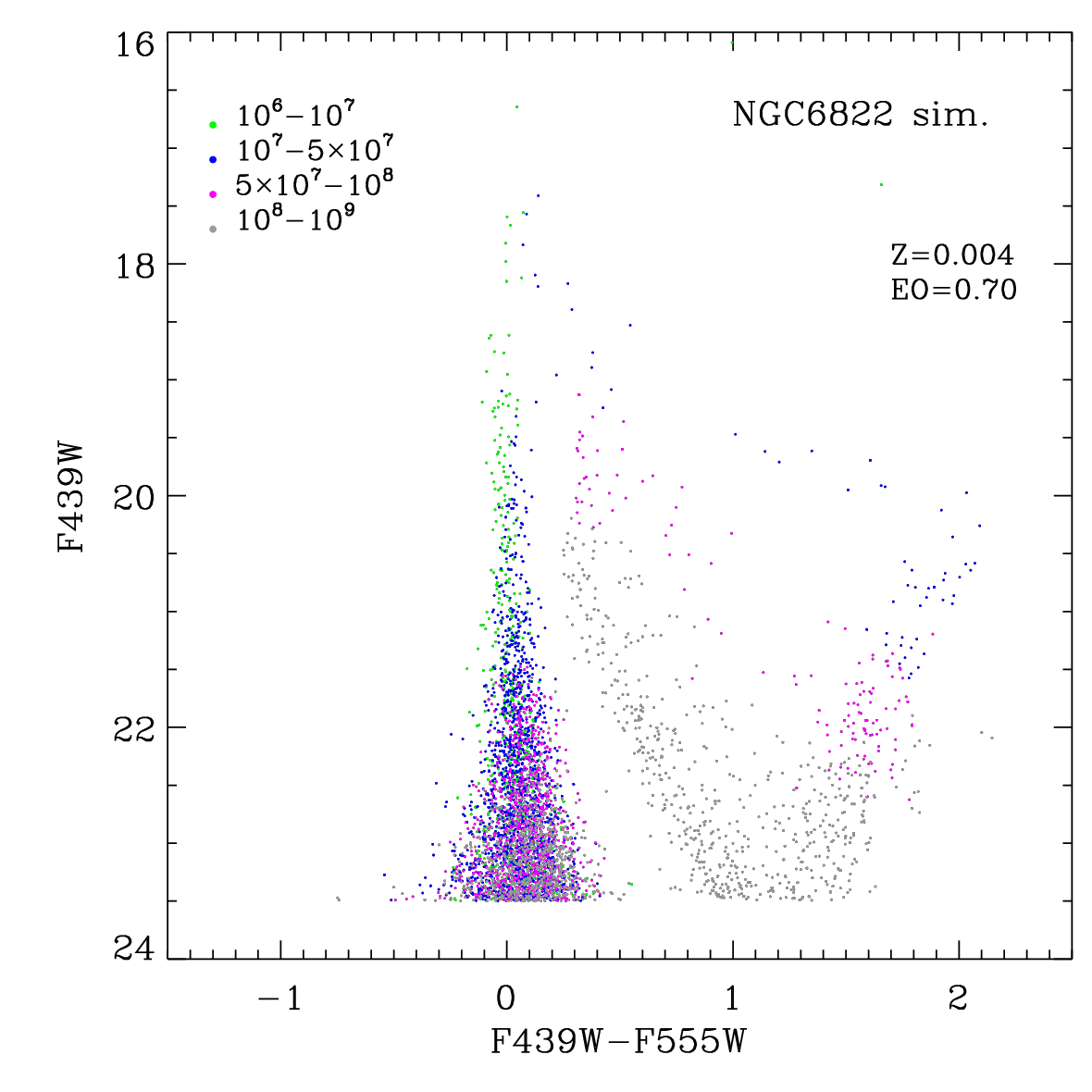}\\
\includegraphics[angle=0,width=0.42\textwidth]{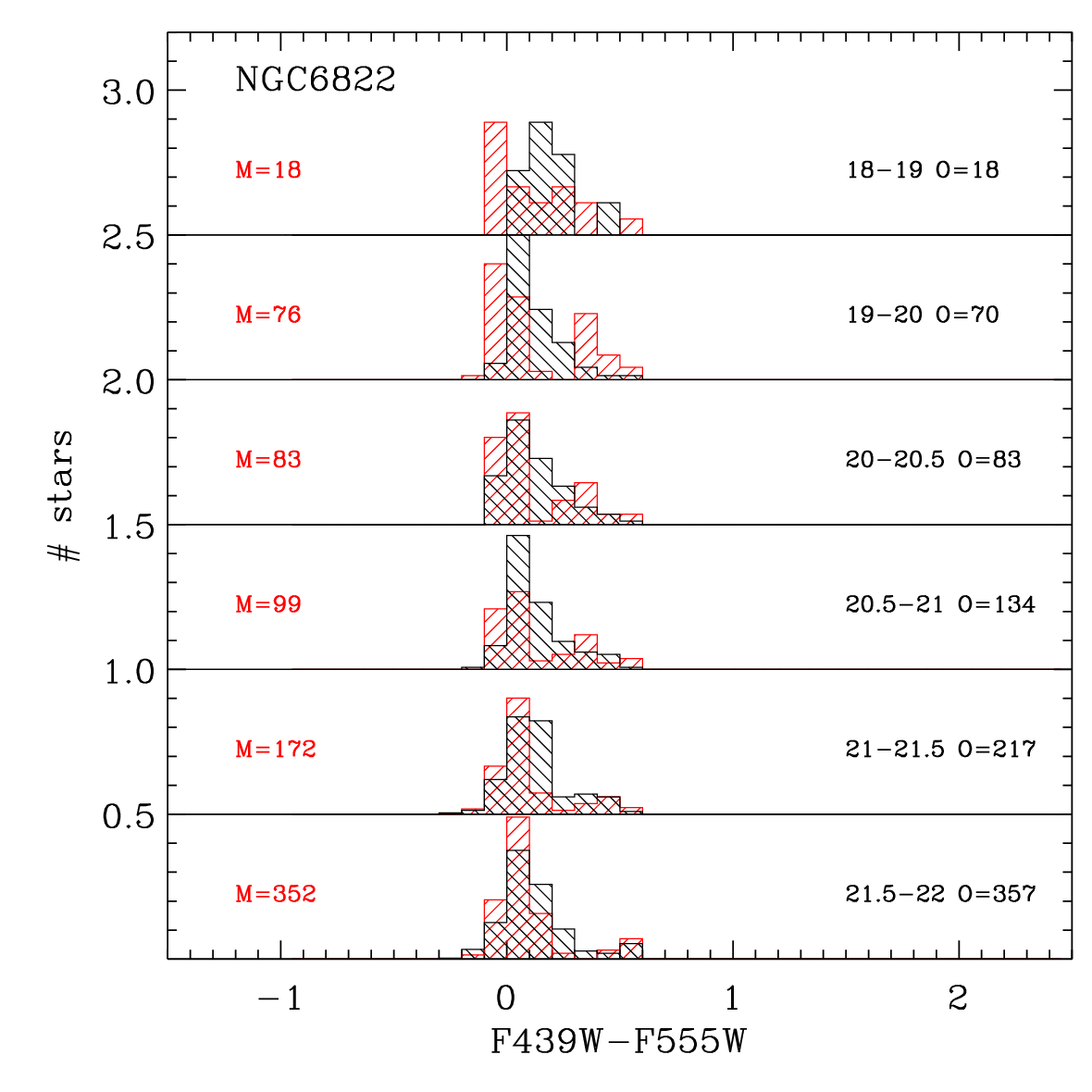}
\caption{Upper panel: Observed color-magnitude diagram of NGC 6822. Middle panel: the best model obtained with {\sl\,PARSEC}~V1.1.
Lower panel: comparison of the observed and modelled colour distributions. \label{NGC6822_SIM}}
\end{figure}
The CMD of  WLM, depicted in the upper panel of Figure \ref{WLM_SIM},
shows a 
broad main sequence with a quite sharp blue edge and a more smooth decline
in the red edge. The red sequence of WLM is poorly defined.

For WLM we adopt a distance modulus (m-M)$_0=24.95$ \citep{Gieren_etal08}.
The best simulation obtained with canonical {\sl\,PARSEC}~V1.1 models is shown in the middle panel of 
the figure.
In this panel we use different colours
for different ages
 and this helps us to interpret 
the main features of the CMD.
The main sequence of WLM appears to span more than five magnitudes in F439W
 but actually,
at  F439W$\lesssim$21, it is mainly  composed by  blue helium-burning stars.  
This is very clear by looking at  
stars in different age ranges.
In the age interval   between 50~${\rm Myr}$ and 100${\rm Myr}$ (red dots), the
main sequence ends at F439W$\sim$22.5 and the blue loop sequence is at least
one magnitude brighter. At ages between 10~${\rm Myr}$ and 50${\rm Myr}$ (dark green)
the main sequence ends at F439W$\sim$21~mag in the simulations. Brighter stars near the main sequence
in this age interval  are also burning central helium in the blue loop.
The brightest star in the simulation is an evolved star of initial mass $M$=40\Msun and age$=$5.3${\rm Myr}$
with luminosity Log(L/L$_\odot$)=5.77 and effective temperature Log(\Teff)=4.24.
This luminous super-giant  star is not in the blue side of the loop as its 
fainter counterparts because,
at such initial masses, helium ignition occurs just after the main sequence. Afterwards
the star slowly moves toward the red super-giant phase while burning central helium (see Figure \ref{fig_HRD_EO_Z001}). Notice that this star
is brighter than the brightest star in the
observed CMD but this is due to the stochastic nature of the stellar birthrate process
at these large masses. Indeed
this is not the most massive star in the simulation, which is instead 
a main sequence star  with initial mass $M$=108\Msun and age$=$1.77${\rm Myr}$,
about 1.3mag fainter.
In any case, it appears clearly
 from the simulation that the right side of
the ``main sequence'' is actually made by
blue helium-burning stars.

Also in the case of WLM  the global luminosity function of this principal sequence is well reproduced with a SFR that increases toward recent times (see Table \ref{tab_SIM_WLM})
The average star formation rate  in the last 100${\rm Myr}$ is $<$SFR$>$=3.7E-3\Msun/${\rm yr}$.

While the global luminosity function of the observed CMD of WLM is quite well reproduced by the simulation, inspection of the colour histograms drawn in the lower panel of Figure \ref{WLM_SIM} 
 shows that the simulated colour distribution across the main sequence does not match the observed one.
There appears a clear excess of stars along the red side of the main sequence  and a lack of very blue stars
which makes the simulated main sequence too broad,
 especially in the magnitude interval
20$\leq$~F439W~$\leq$22.
Note 
that the models adopted for WLM have the same metallicity ($Z$=0.001) as
those used for the simulation of
 Sextans A.
In the CMD of Sextans A the main sequence and the blue loop sequence were clearly separated
while, in the case of WLM, this separation is not so evident because of
the larger errors associated with the observations.
The apparent excess of red stars is due to the blue helium-burning sequence,
therefore
we conclude that also
in the case of WLM there is a
discrepancy concerning these stars.
\begin{table}
\begin{center}
\caption{Parameters of the CMD simulations of NGC6822\label{tab_SIM_NGC6822}}
\begin{tabular}{llllll}
\hline\hline
EO & T$_i[{\rm yr}]$ &T$_f[{\rm yr}]$ & $\tau[{\rm yr}]$ & X$_{IMF}$ & $<$SFR$>[{\rm M_\odot/yr}]$ \\
\hline
0.7 &  1E6 &1E9 &-2E8 & 2.35 &  3.5E-3\\
2   &  1E6 &1E9 &-2E8 & 2.35 &  3.7E-3\\
4   &  1E6 &1E9 &-2E8 & 2.55 &  4.3E-3\\
\hline
\end{tabular}
\end{center}
\end{table}

\begin{figure*}
\includegraphics[angle=0,width=0.42\textwidth]{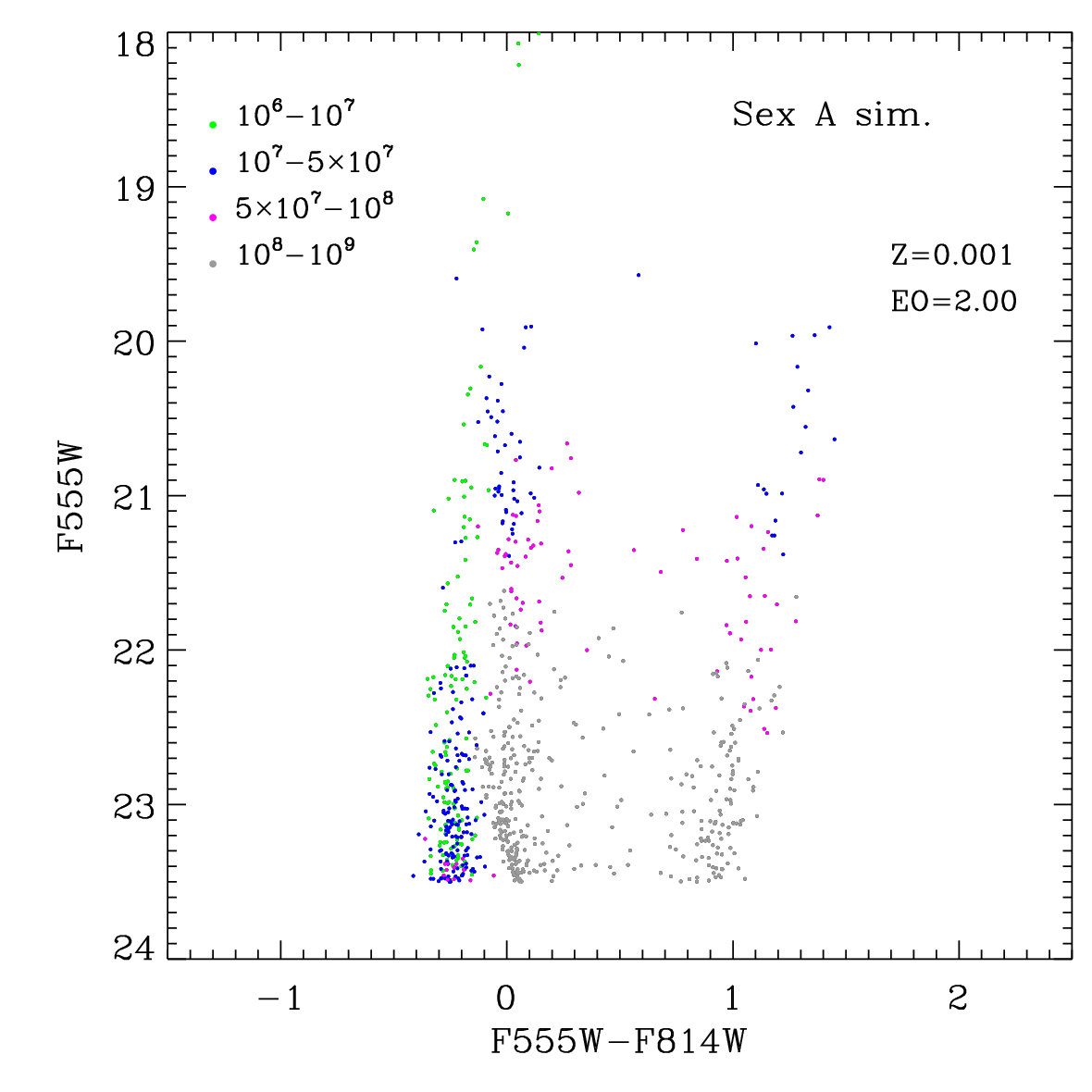}
\includegraphics[angle=0,width=0.42\textwidth]{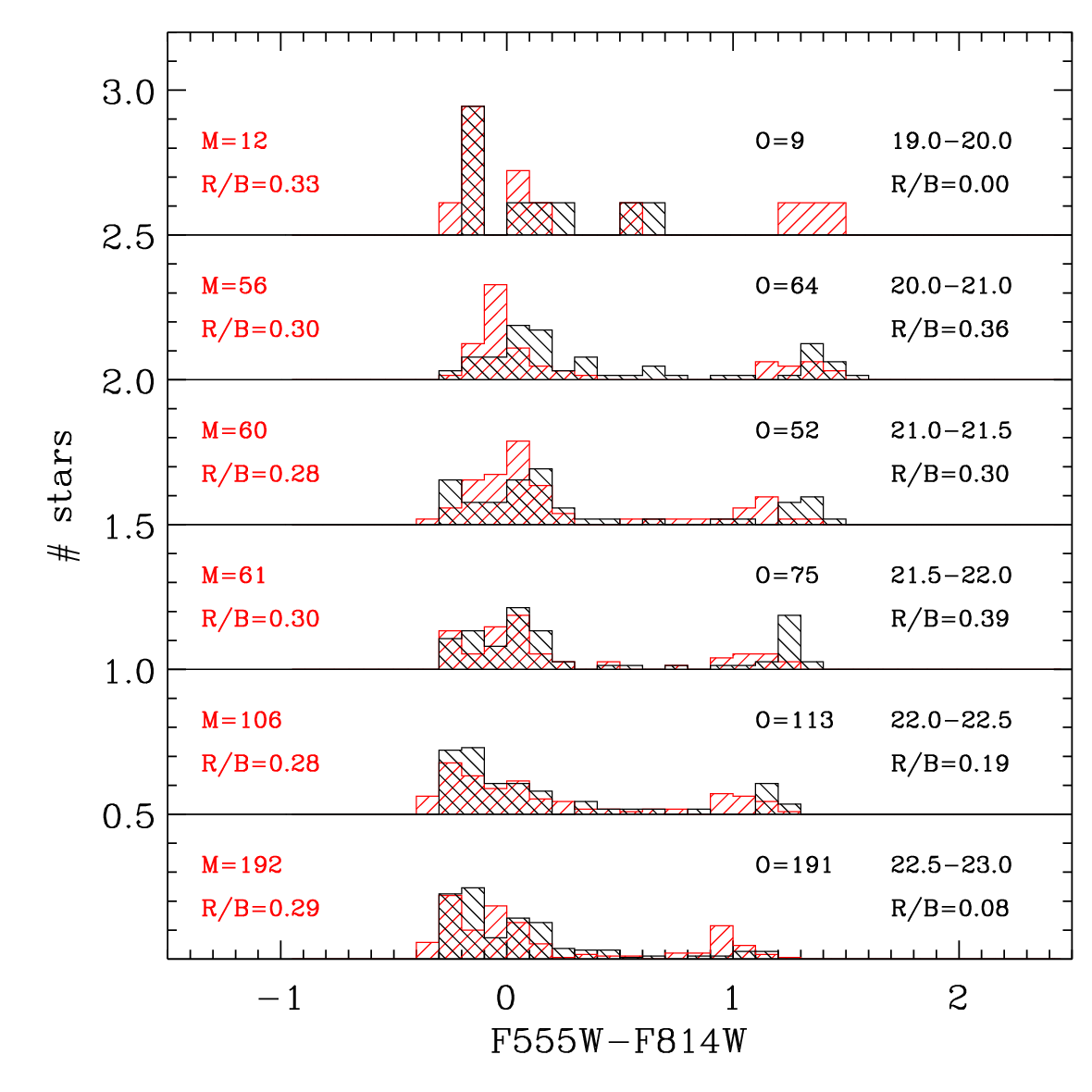}\\
\includegraphics[angle=0,width=0.42\textwidth]{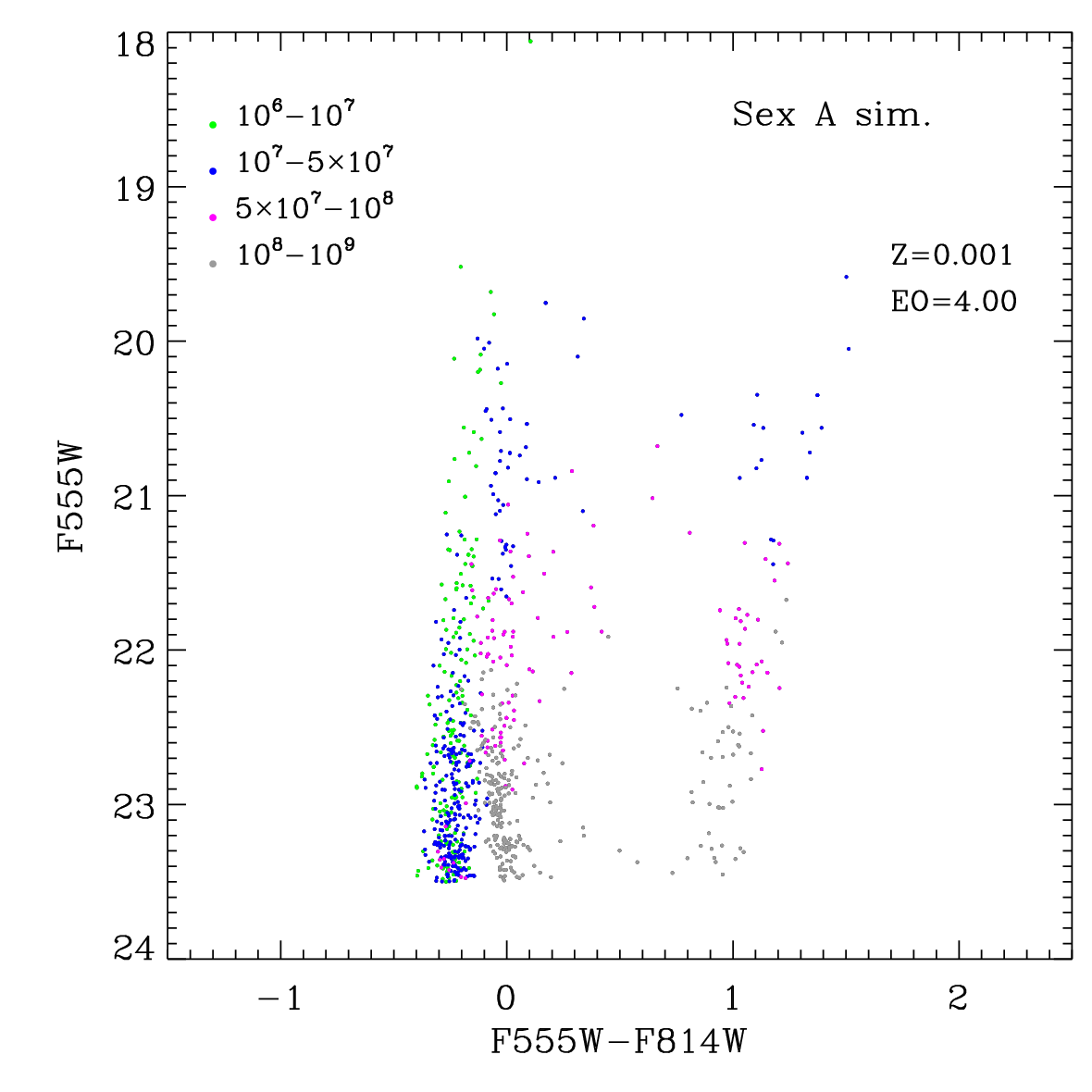}
\includegraphics[angle=0,width=0.42\textwidth]{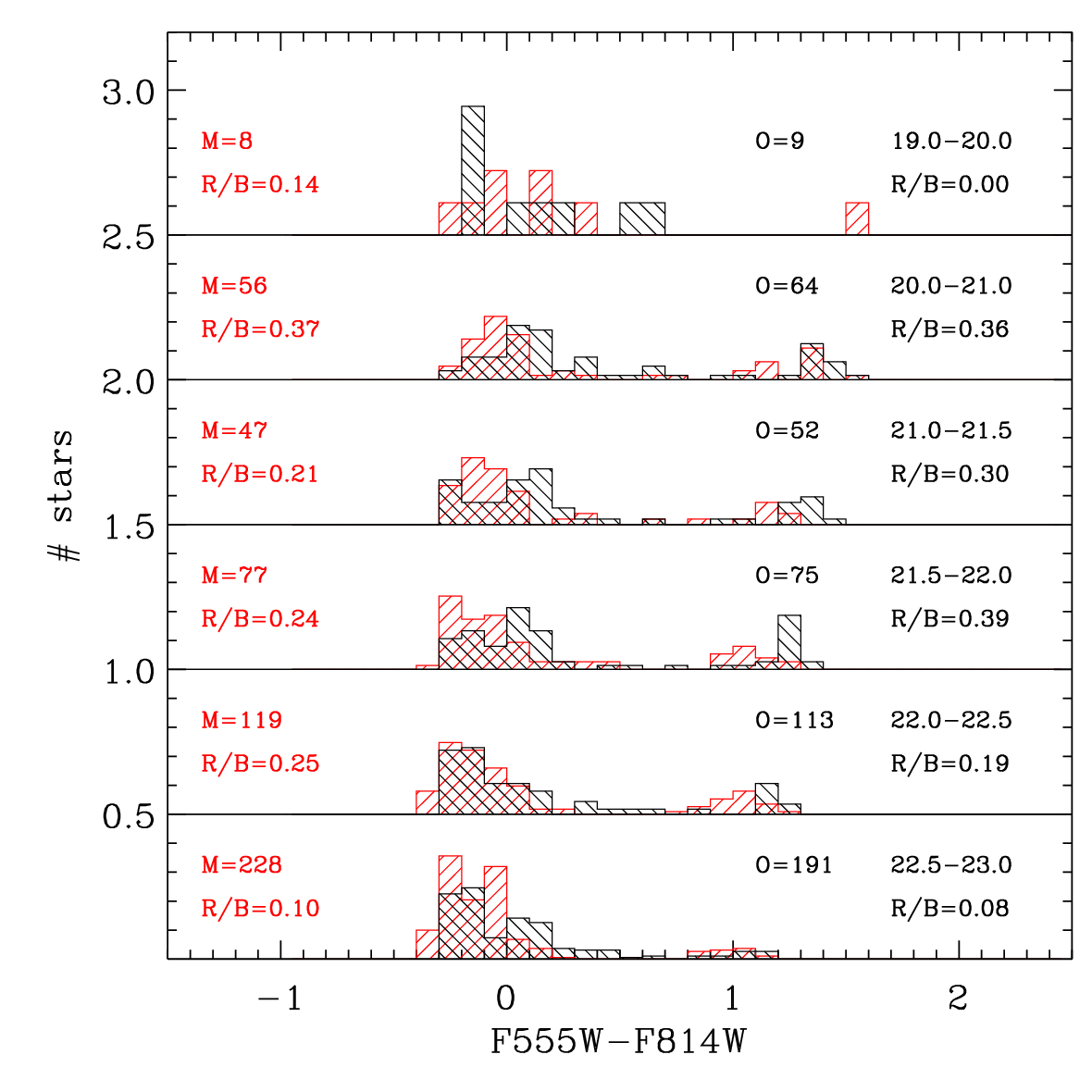}
\caption{Upper panels: the best simulated CMD of Sextans~A (left) and the corresponding star colour distribution (right), obtained with {\sl\,PARSEC} increasing the envelope overshooting to EO=2\HP.
Lower panels: the same as in the upper panels but for a larger envelope overshooting, EO=4\HP.\label{SEX_A_SIM_EO2}}
\end{figure*}
\subsection{NGC6822}
The observed CMD of  NGC6822 is shown in the upper panel of Figure \ref{NGC6822_SIM}.
With a distance modulus (m-M)$_0=23.31$ \citep{Gieren_etal06},
NGC6822 is the nearest of the three galaxies analysed here and its observed blue sequence spans more than seven magnitudes.
In the CMD we recognize also a yellow sequence, around F439W-F555W$\sim$0.8/0.9, and another
redder  sequence. It is important to clarify the origin of the yellow sequence, because 
the  metallicity of NGC6822, $Z$$~\sim$0.004, is definitely
higher  than that of the other two galaxies, 
and this sequence could trace the position of the {\it blue} loops of relatively metal-rich galaxies. However, our
 TRILEGAL simulations, and comparison of the CMD among different galaxies,
indicate
a large contribution from Milky Way foreground stars near the position of the middle sequence.
Actually there is some mismatch between
the predicted position of the MW stars and the observed CMD. The predicted MW foreground stars
not only populate 
  the region of the observed yellow sequence (mainly thin disk stars)
but also a region a few tenths 
of magnitude bluer. 
Furthermore, significant  contamination by thin disk stars
extends 
 also brighter than the observed F439W$=$18mag, where only a few stars are seen in the CMD.
Thus, due to the significant contamination by Milky Way stars, we exclude the yellow sequence from our analysis.
Taking into account that many predicted foreground stars
are slightly bluer than the observed yellow sequence,  we
exclude from further analysis
all stars with observed colour redder than F439W-F555W$=$0.6. Below this threshold the contamination
becomes negligible and we  may safely use the well populated blue sequence of NGC6822 to
check the models.

The best simulation obtained with canonical {\sl\,PARSEC}~V1.1 models is shown in the middle panel of Figure \ref{NGC6822_SIM}.
The  parameters of the stellar birthrate for this simulation are shown in Table \ref{tab_SIM_NGC6822}.
The star formation rate increases toward more recent times  with an average value in the last 100~${\rm Myr}$  of $<$SFR$>$=3.5E-3\Msun/${\rm yr}$.
Note that the observed CMD includes seven HST fields, which cover a fraction of the galaxy
\citep[$\sim$0.73kpc$^2$, see][Table 1 and Figure 1]{Bianchi_etal12}. Because
star formation is very patchy in NGC~6822, and on the other hand the HST pointings used here
 targeted star-forming sites, this SFR cannot be simply scaled to the whole
galaxy (the two most conspicuous
star-forming regions, including Hubble~V and Hubble~X, and an outer older region, were observed by \citealt{Bianchi_etal01} and \citealt{Bianchi_etal06}.

We first notice that, from the {\sl\,PARSEC} 
simulation, very few stars are expected to lie
in the region occupied by the yellow sequence, corroborating the idea that this sequence is
due to strong contamination by MW foreground stars.
The maximum mass in the simulation is
$M$=75\Msun at an age of 2.1~${\rm Myr}$, while
the brightest star has an initial mass $M$=72\Msun and an age of 3.6~${\rm Myr}$
The simulation suggests the presence of a sequence of blue helium burning stars
running parallel to the main sequence. However in the simulated diagram this sequence appears clearly detached from the main sequence while in the observed CMD
it appears a continuation of the main sequence.
This appears more clearly in the comparison of the colour distributions in the lower panel of
Figure \ref{NGC6822_SIM}. Contrary to the observed distributions, the predicted ones
 in the range of magnitudes between 19$\leq$~F439W~$\leq$21.5 are bimodal.
\begin{figure}
\includegraphics[angle=0,width=0.4\textwidth]{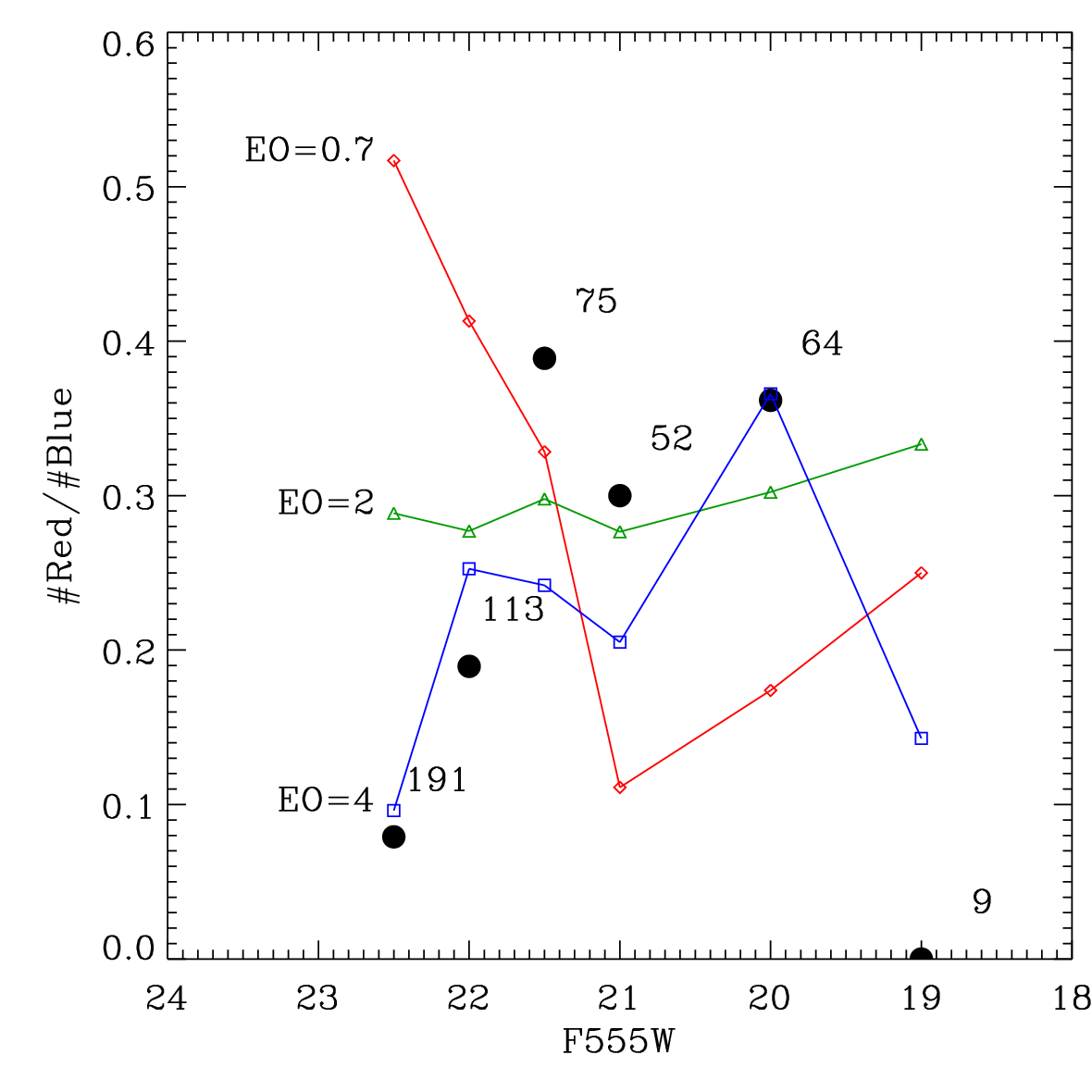}\\
\caption{Observed (large filled 
dots) and predicted (coloured lines with symbols) ratios between the number of stars redder and bluer than  F555W-F814W $=$0.6 as a function of the apparent magnitude, in the CMDs of Sextans A. The numbers indicate the
total number of stars observed in each magnitude bin.
The predicted ratios are labelled with the envelope overshooting values.
\label{fig:SEXTANSA_RB}}
\end{figure}
\section{Results with enhanced envelope overshooting models}\label{simhrenvov}
The above comparisons indicate
that the adopted stellar evolutionary models cannot reproduce well
the observed CMD:
even accounting for
errors and reddening
which tend to smear out the colour distribution, there remain a visible gap between
the main sequence and the blue loop helium-burning sequence.
This gap is not observed or much less evident in the data and, at the same time,
the colour distribution of the models is wider than that observed.
It is worth noticing that
adopting a lower metal content we could
remove the discrepancy
between
modelled and observed CMD, but
this would require a metallicity value quite discrepant
from the one determined  from spectroscopic observations.
Moreover, the value adopted in the simulations of WLM and Sextans A, $Z$=0.001, lies in between the
ones derived observationally for the two galaxies, indicating that
the origin of the discrepancy is not due the adoption of a too high metallicity.
Given the inability of reproducing the observed blue loops by means of models with canonical envelope overshooting
we have  repeated the simulations adopting  models with enhanced envelope overshooting.

The best simulations of Sextans A obtained with models with enhanced envelope overshooting  are shown in the upper and lower 
panels of Figure \ref{SEX_A_SIM_EO2}
for the cases of EO=2\HP and EO=4\HP, respectively.
As already discussed in section \ref{hrenvov},
models with enhanced envelope overshooting perform more extended blue loops and
this is visible in the corresponding simulations.
Adopting EO=2\HP produces already 
a reasonably good fit to the observed CMD
of Sextans A. The bimodal behaviour noticed in Figure \ref{SEX_A_SIM} at fainter magnitudes disappears and
the excess of stars in the red side of the main sequence becomes significantly less
pronounced. With EO=4\HP the loops are even more extended
and the predicted main sequence becomes even narrower than the observed one.
The star formation rate is not significantly different from the one derived
with the standard {\sl\,PARSEC}~V1.1 models. It increases by about 16\% using the
models with EO=2\HP  and by another 10\% if one consider the models with
EO=4\HP. The latter increase is in part due to the decrease of
the star formation time-scale adopted for the best fit (Table \ref{tab_SIM_SEXA}).

The observed CMD of Sextans A is so well populated along the different sequences,
that we may gain
 some insight on the properties of the different mixing models
by analysing the number ratios between red
(F555W-F814W $>$0.6 and F555W-F814W $<$(31.2-F555W )/7)  and blue stars
(F555W-F814W $\le$0.6).
The observed number ratios are plotted against the apparent magnitude in
Figure \ref{fig:SEXTANSA_RB} with large filled dots.
The number ratios predicted by adopting the
different values of the overshooting scale are plotted with different symbols and colours and are labelled
by the adopted overshooting scale.
The size of the envelope overshooting affects not only the extension of the loop
but also the relative lifetimes in the blue and red sides.
In the models with larger envelope overshooting,
the loops begin at earlier  times during central helium burning
(Figure \ref{Psi_laut}), and
 the relative number of red to blue stars
decreases. The observed  number ratios of stars in the red and blue side
of the loops have been already used
to check the performance of stellar evolution models,
because they 
concern post main-sequence phases, hence they depend almost exclusively
on the lifetime ratios of the stars in the corresponding  phases \citep[e.g.][]{Dohm_etal02}.
We cannot perform the same comparison here because we cannot disentangle
the blue side of the loop and the main sequence. 
Indeed in Figure \ref{fig:SEXTANSA_RB}
we plot the ratios between the number of red stars and that of blue stars
which include also the main sequence stars. Thus, these ratios
are not strictly related to the above lifetime ratios not only because
they include the main sequence phase but also because,
by selecting them in bins of constant magnitude, we are including stars with different masses.
Nevertheless, the exercise is meaningful because we are simulating the
real evolution of these stars in the CMD.
Indeed the figure shows that, for higher 
envelope overshooting, the R/B ratio decreases significantly, because the red side of the loops become less and less populated.

\begin{figure*}
\includegraphics[angle=0,width=0.42\textwidth]{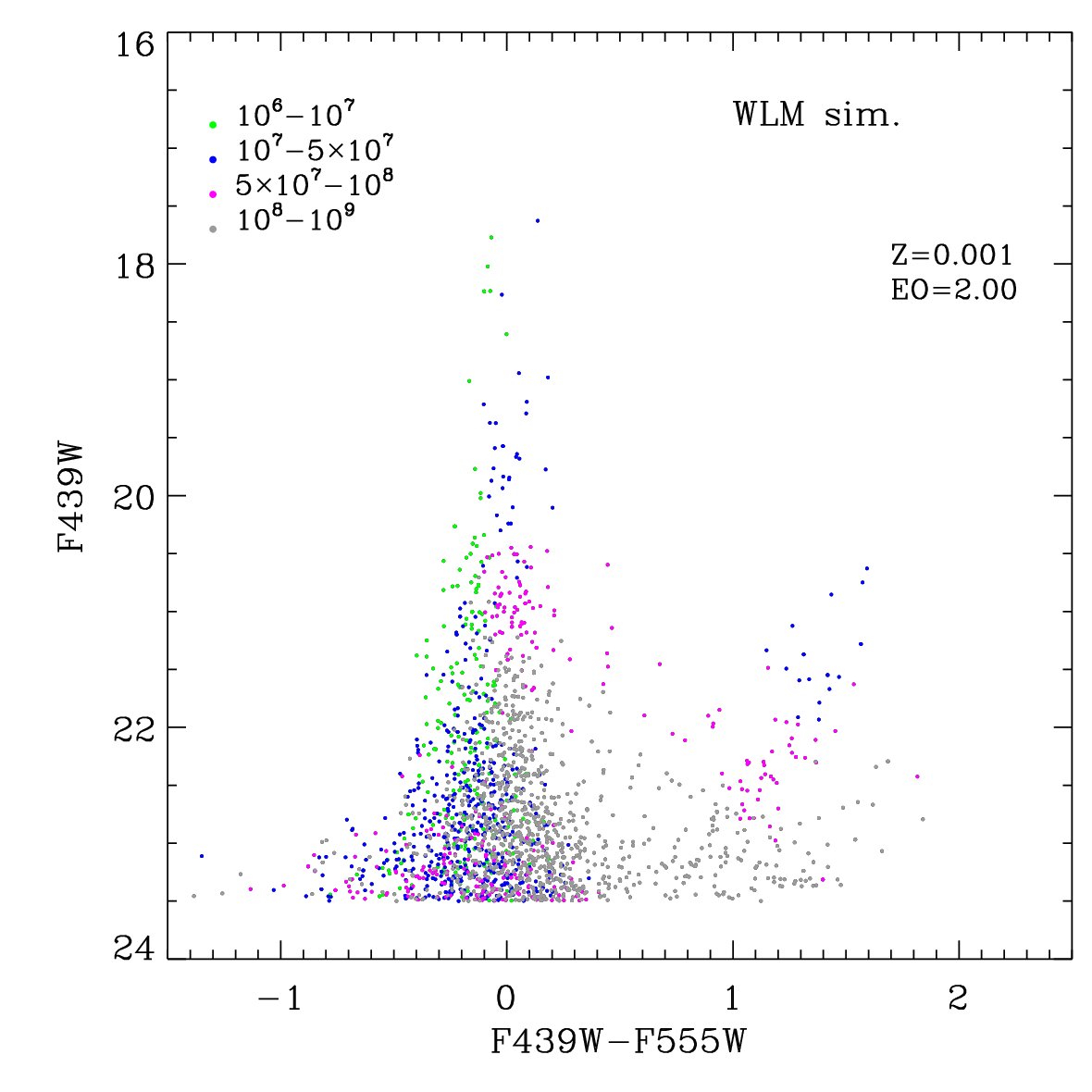}
\includegraphics[angle=0,width=0.42\textwidth]{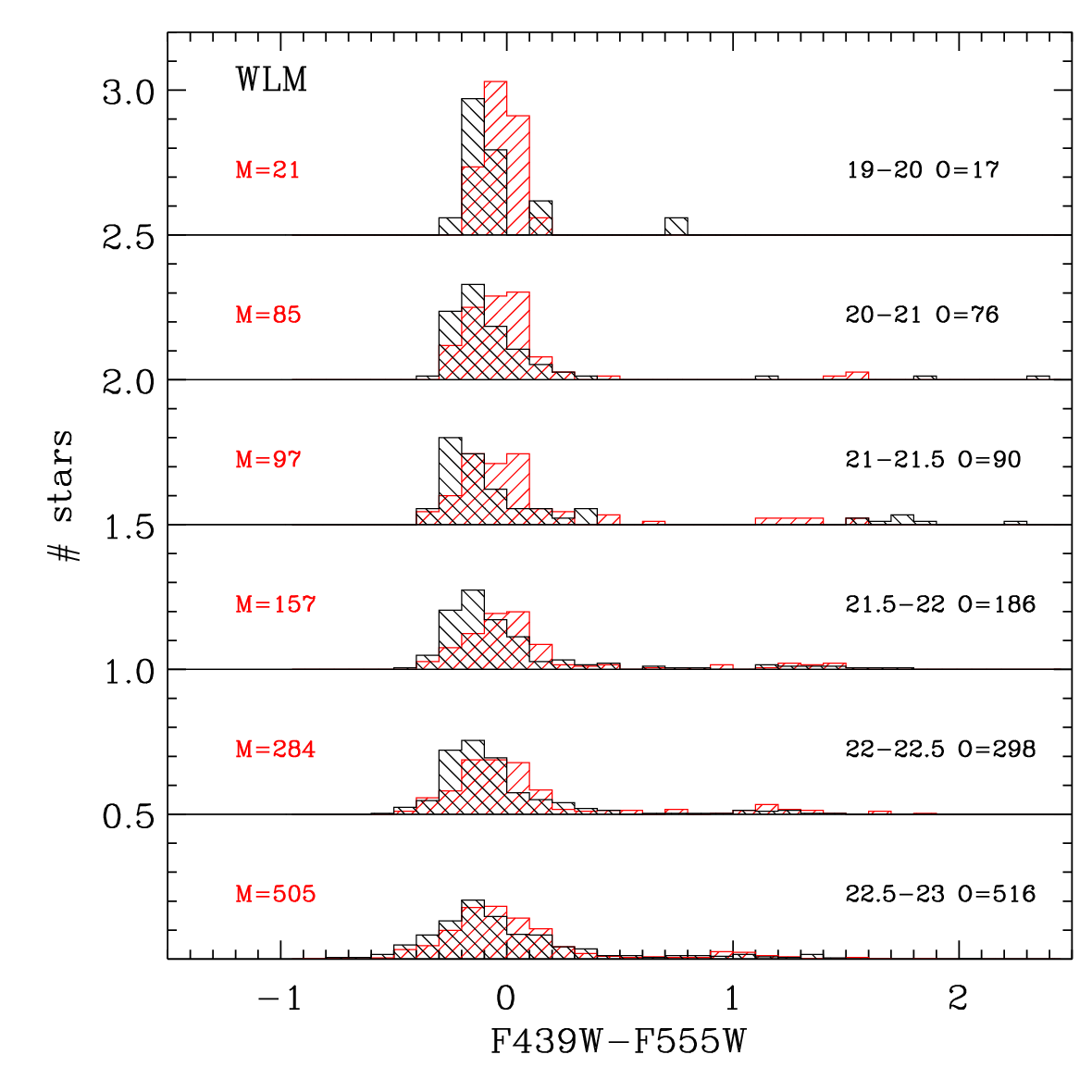}\\
\includegraphics[angle=0,width=0.42\textwidth]{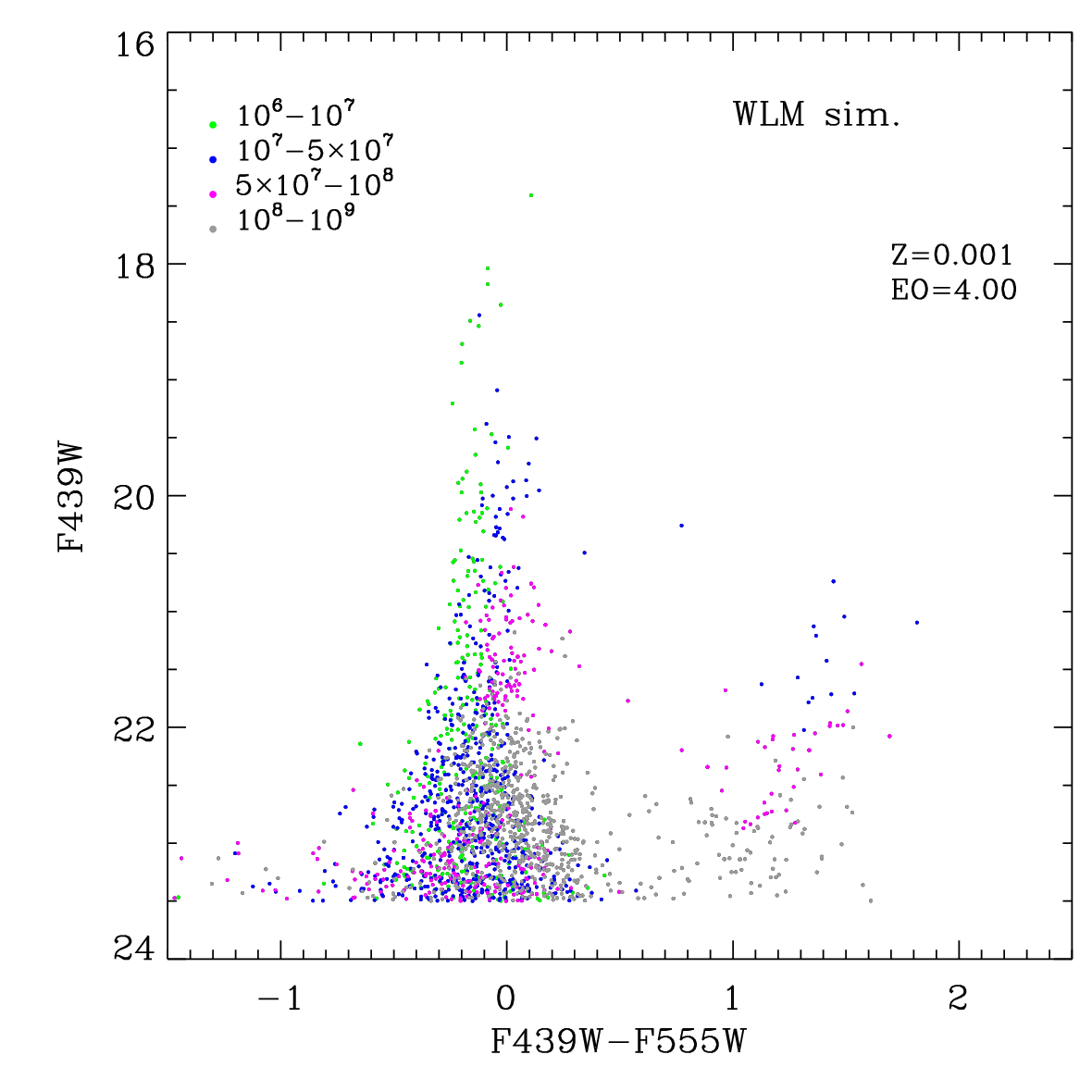}
\includegraphics[angle=0,width=0.42\textwidth]{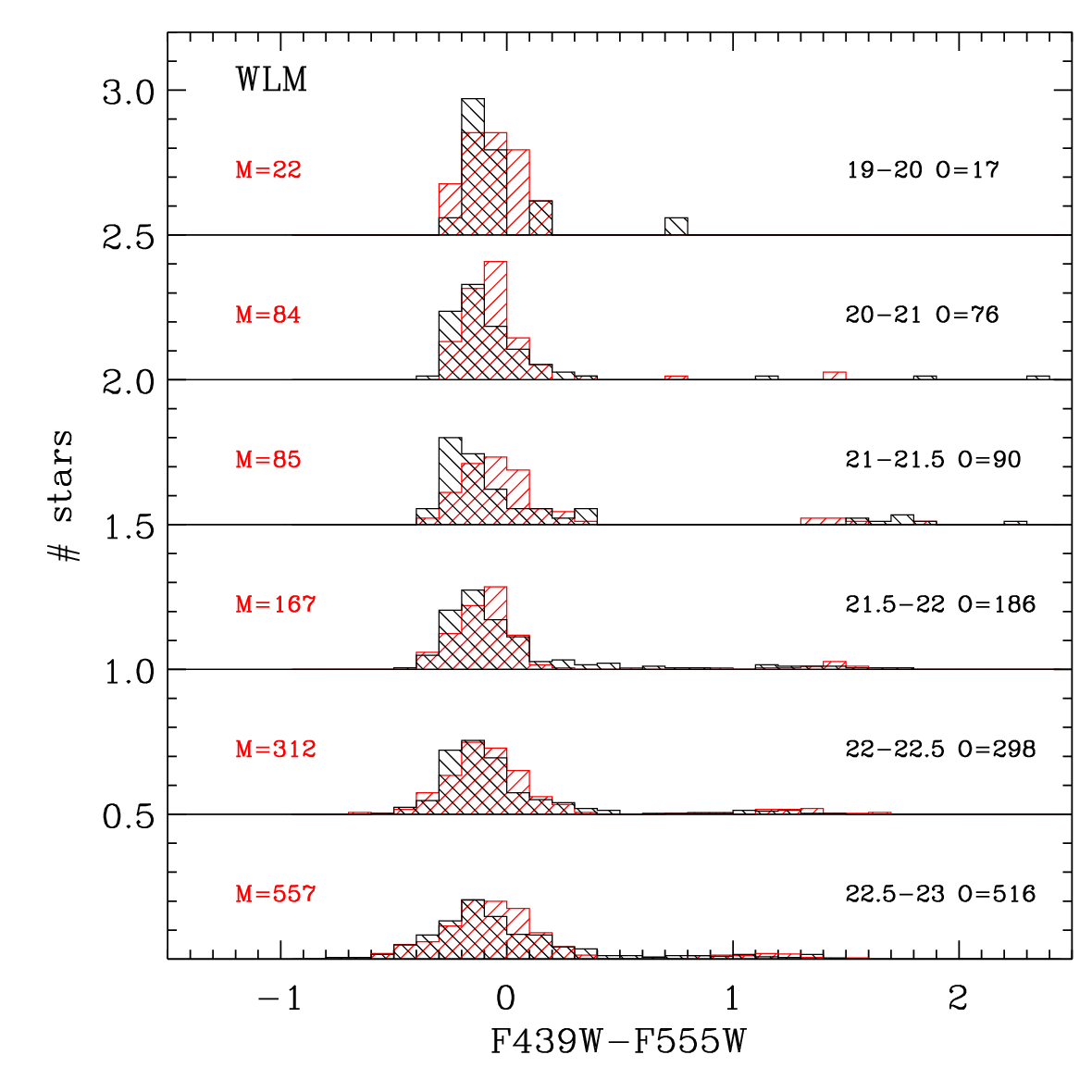}
\caption{Upper panels: the best simulated CMD of WLM (left)and the corresponding star colour distribution (right), obtained with {\sl\,PARSEC} increasing the envelope overshooting to EO=2\HP.
Lower panels: the same as in the upper panels but for a larger envelope overshooting EO=4\HP.
\label{WLM_SIM_EO2}}
\end{figure*}

The best simulation of WLM obtained with these new models is
shown in
Figure \ref{WLM_SIM_EO2}
for the cases of EO=2\HP and
 EO=4\HP.
Again in the case
of WLM the simulations performed with enhanced envelope overshooting models
better agree with the observations. The relative excess of stars in the red side
of the main sequence decreases significantly with EO=2\HP and almost disappears
with EO=4\HP. Compared with
the standard case of
EO=0.7\HP,
the average star formation rate in the new best-match models
is about 30\% lower.
This is due to the different IMF slope adopted in
the first case, x=2.65, with respect to the one adopted in the latter cases
x=2.35.
\begin{figure*}
\includegraphics[angle=0,width=0.42\textwidth]{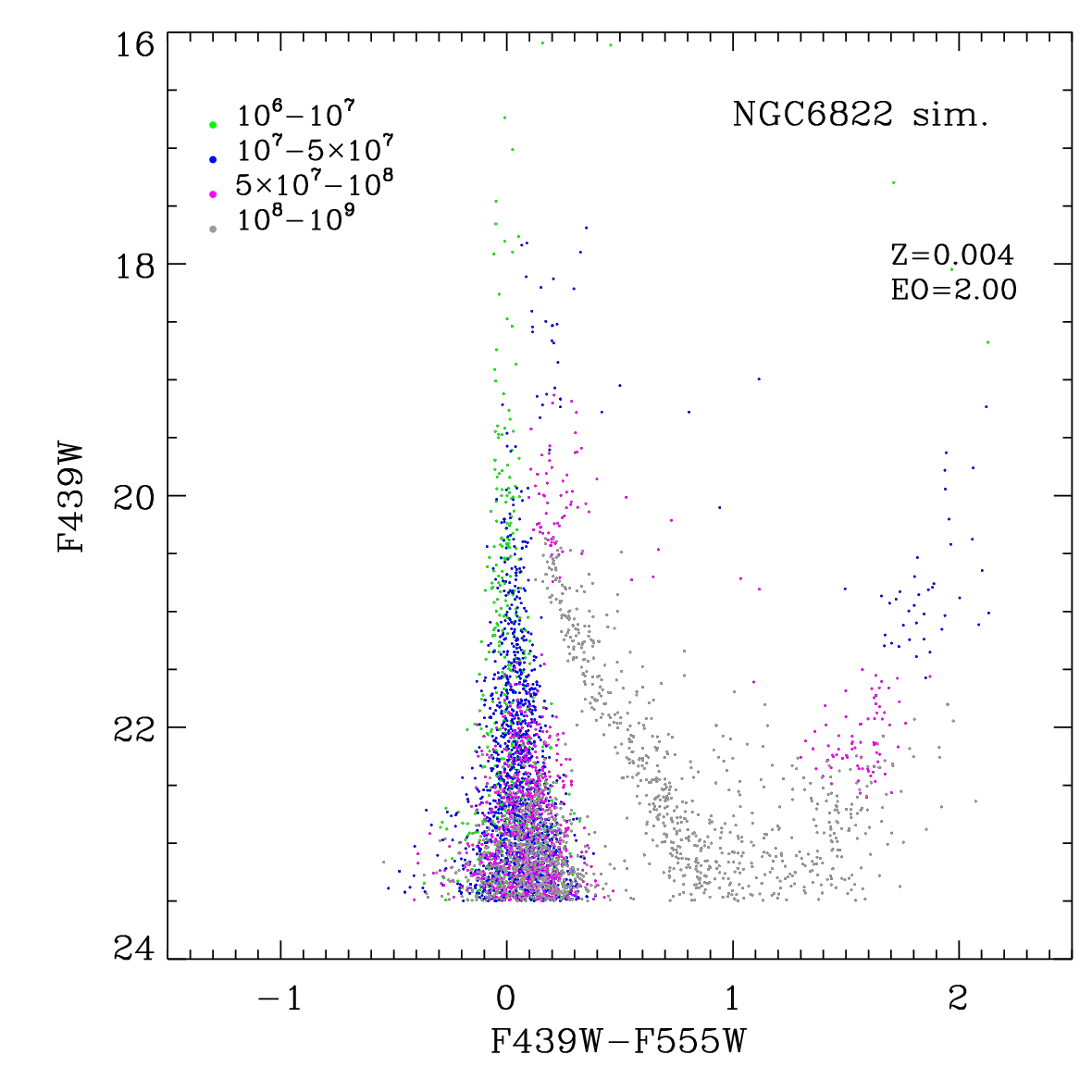}
\includegraphics[angle=0,width=0.42\textwidth]{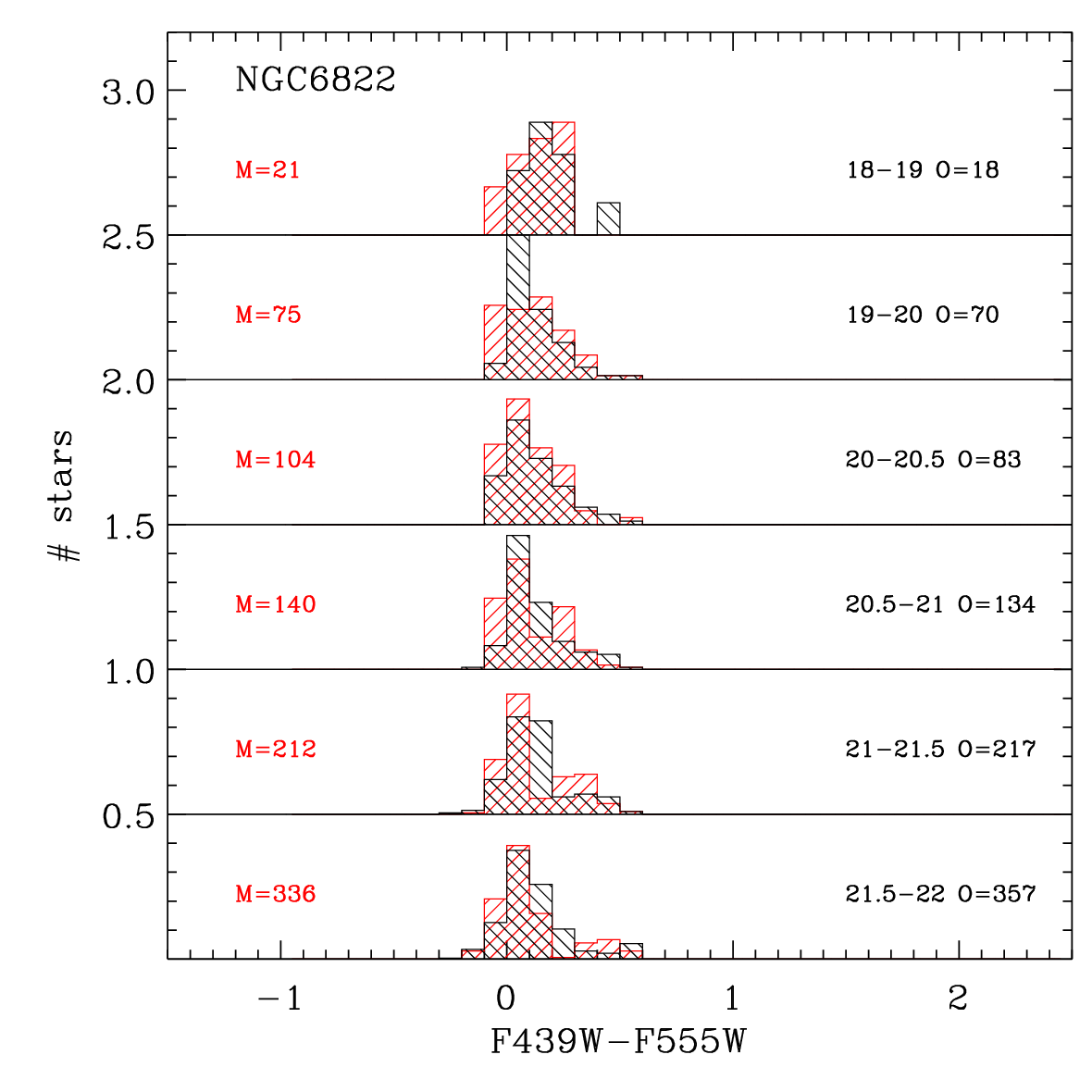}\\
\includegraphics[angle=0,width=0.42\textwidth]{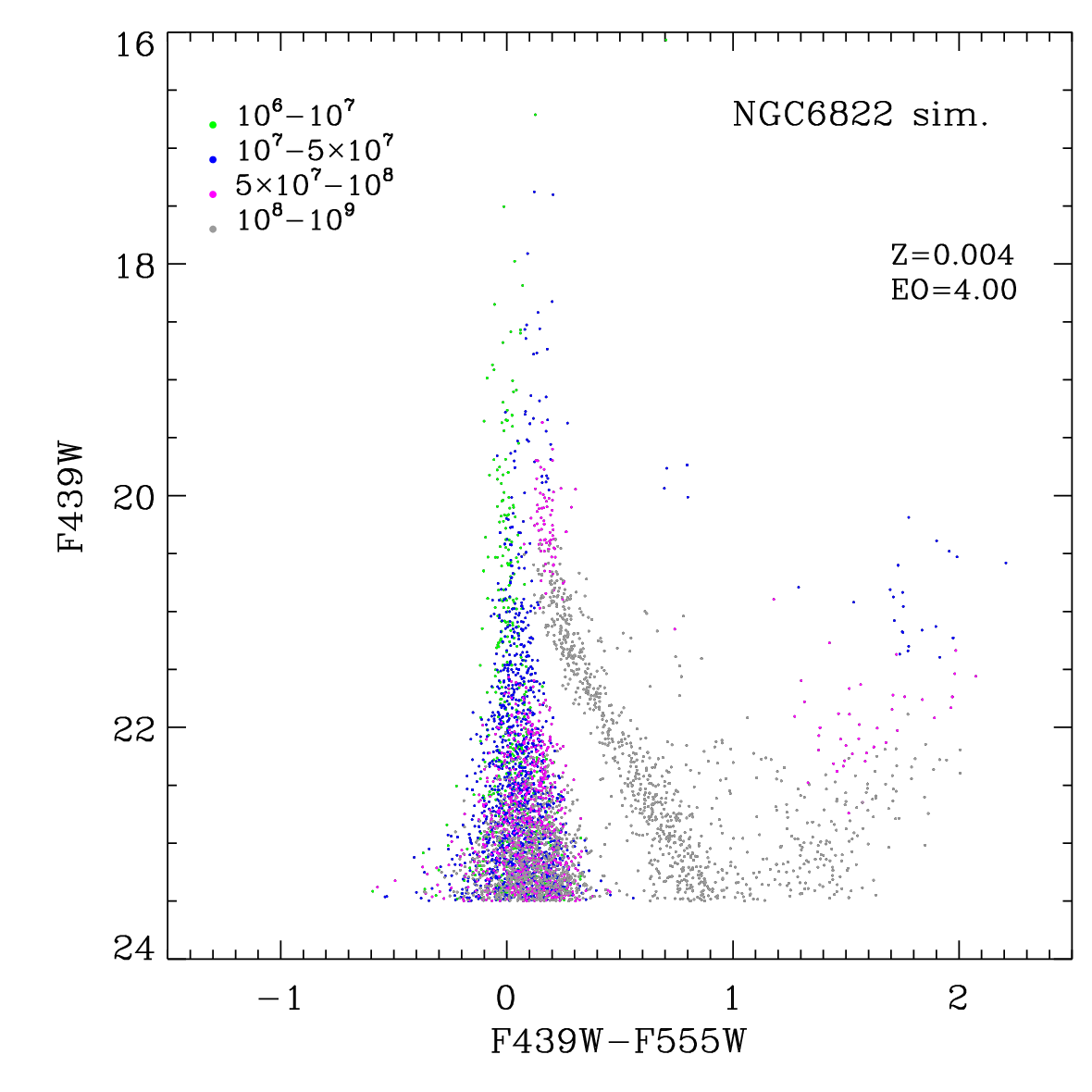}
\includegraphics[angle=0,width=0.42\textwidth]{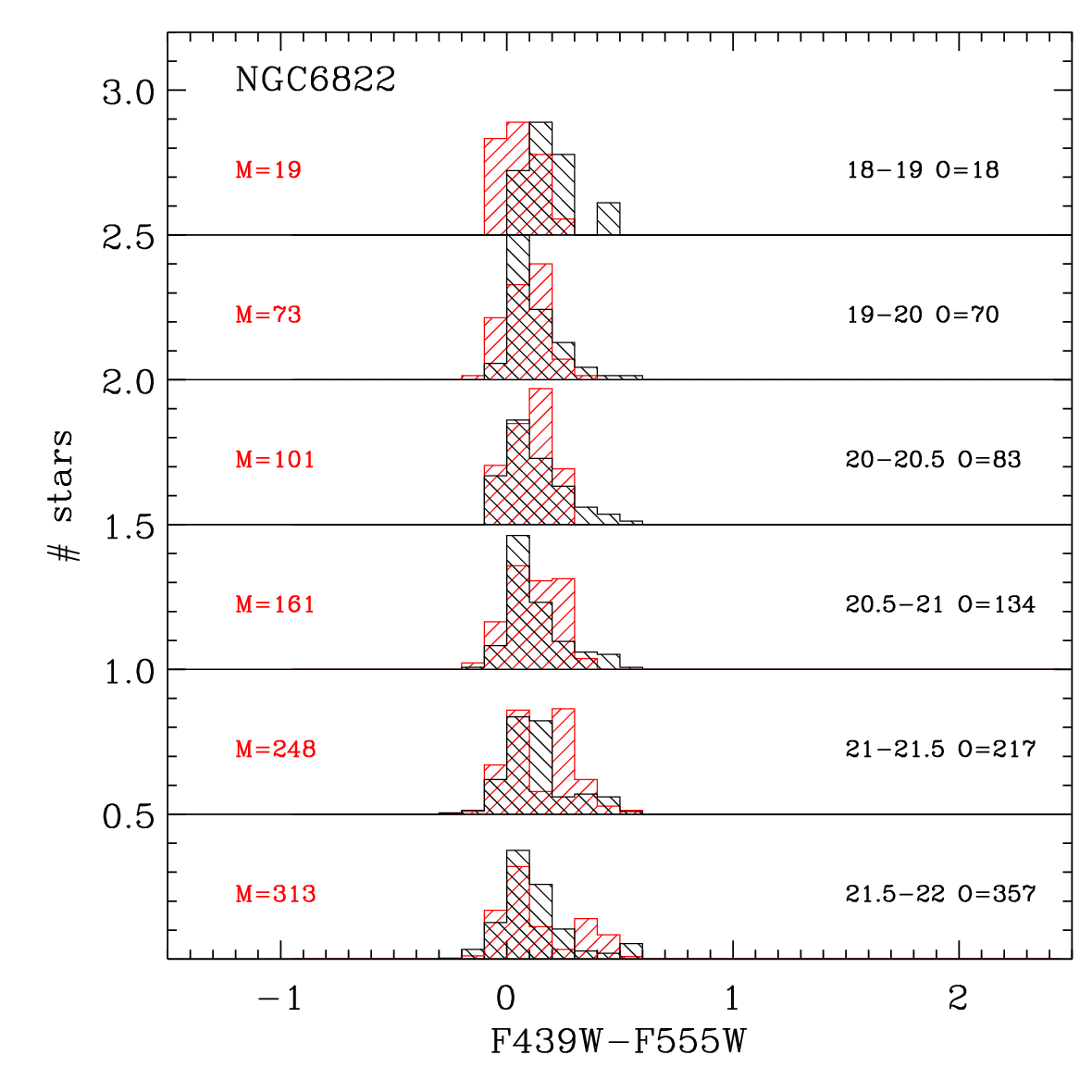}
\caption{
Upper panels: the best simulated CMD of NGC~6822 (left)and the corresponding star colour distribution (right), obtained with {\sl\,PARSEC} increasing the envelope overshooting to EO=2\HP.
Lower panels: the same as in the upper panels but for a larger envelope overshooting EO=4\HP.\label{NGC6822_SIM_EO2}}
\end{figure*}
For NGC6822, the best simulations
obtained with models with enhanced envelope overshooting  are shown in the upper and lower panels of Figure \ref{NGC6822_SIM_EO2},
for the case  EO=2\HP and EO=4\HP, respectively.
Even in this case the morphology of the blue sequence is
better reproduced.

\section{Conclusions}\label{sec:conclusion}
With the aim of extending the new library of stellar evolutionary tracks and isochrones computed with {\sl\,PARSEC},
we begin the calculation of new evolutionary tracks of massive stars.
We adopt the same input physics used in the {\sl\,PARSEC}~V1.1  published version
of low and intermediate mass stars, the only difference being that for masses
$M\geq$14\Msun we include mass-loss.
The new calculations supersede the old stellar evolution tracks of massive stars
\citep{Bertelli_etal94},
so that we now provide updated and homogeneous
sets of evolutionary tracks
from very low ($M$=0.1\Msun) to very massive ($M$=350\Msun) stars,
from the pre-main sequence to the beginning of central carbon burning. 

In this paper, we presented
new evolutionary tracks of low metallicity,
$Z$=0.001 and $Z$=0.004. We also performed
comparison
of the new tracks with
observed CMDs of star forming regions in three selected nearby metal-poor dwarf irregular galaxies, Sextans A, WLM and NCG6822.
These galaxies are dominated by the last ongoing star-burst and the existing
HST CMDs are a useful workbench  for testing the overall performance of stellar evolution tracks in the domain of intermediate and massive stars.
The contamination by Milky Way dwarf stars,
estimated with TRILEGAL simulations,
is negligible in Sextans A and WLM,
while
for NCG6822 it is significant at colours F439W-F555W $\geq$0.6:
for this galaxy we used
 only bluer stars,
which include the main sequence and the blue helium-burning sequence.

Since the new massive star tracks have not yet  been implemented in the popular CMD simulators
(e.g. TRILEGAL) we built
our own CMD simulator, where we specify
the IMF and SFR laws, the  metallicity, the photometric errors, attenuation and binary effects.
In the simulations, we fix the  metallicity to the value which is the nearest
to that found in literature for the youngest populations of the galaxy, either stars or gas. This is  important because our aim is to check the model results by using the morphology of the
 CMD, in particular of the BHeBS, which
 is known to depend significantly on the metallicity.
The photometric errors are taken
from the published
photometry catalogues and are reproduced  statistically
 as a function of the apparent magnitude.
As for the attenuation by IS dust,
we exploit the advantage that it has been individually
derived for a large sub-sample of stars, by means of a star by star multiband spectro-photometric analysis \citep{Bianchi_etal12}.
In all the three galaxies, the attenuation is characterized by an average value that rises  with the intrinsic luminosity  for the hot stars and by a significant dispersion, which is consistent with
an age selective extinction together with local variations
of the dust content.
Only by including such a dispersion  and trend with luminosity we may reproduce the morphology of the different sequences seen in the CMD diagram, without a real need to use a dispersion in metallicity.
This is especially true for Sextans A where the photometric errors are very small while, in the case of NGC~6822, there is still room for a dispersion in metallicity,  which indeed has  been observed in
this galaxy, and is not surprising given its patchy star-formation episodes.
We check also the effects of non physical binaries and find that
they are not  very significant so that we do not discuss them in the paper.

For Sextans A and WLM we adopted  a metallicity $Z$=0.001, which is an upper limit  for the former and a lower limit for the latter.
In both galaxies the location of the main sequence is well reproduced by the new tracks.
Sextans A  shows a well populated red sequence which is also fairly well reproduced by the new models. In Sextans A the main sequence and BHeBS form two parallel vertical sequences while
in WLM they form a broad blue sequence.
All models generally reproduce
 these sequences but,
a quantitative comparison of
the colour distribution
 in different magnitude  bins
shows
that, for models with canonical overshooting,
the simulated distributions  are broader than the observed ones and generally skewed toward redder colours, in both galaxies.
A lower metallicity, which would render the blue loops more extended,
would be discrepant with measured values, therefore
we interpret this mismatch as evidence that the BHeBS predicted by the models are not hot enough.

For NGC6822 we adopt a metallicity of $Z$=0.004. 
As for the previous galaxies,
the fit of the main sequence is fairly good, but at this metallicity the problem of the blue loops is even more exacerbated. The models computed with {\sl\,PARSEC}~V1.1  (canonical overshooting) show a shortening of the loops
above \logL$\sim$3.5 (Figure \ref{fig_HRD_EO_Z004}) and the corresponding simulation shows a blue sequence with  bimodal colour distribution which is not evident in the observed CMD.
This corroborates the notion that even at $Z$=0.004
the BHeBS predicted by the models are not hot enough and that,
in order to reproduce the observed morphology of the CMD with the {\sl\,PARSEC}~V1.1 prescriptions,
a metallicity
much lower than the measured values would be required.

We show that this discrepancy  is overcome by extending the overshooting at the base of the convective envelope. The simulations made with the enhanced EO models, better reproduce  the
observed CMDs for all the three galaxies.
For Sextans A the comparison can be extended to the
numebr ratios between stars in the red side of the loop
and blue stars (including also main sequence stars). The comparison with
the observed ratios strongly support the largest value, EO=4\HP.
Thus there is evidence that an EO larger than that adopted in {\sl\,PARSEC}~V1.1 should be preferred.
Apart from the morphology (and corresponding luminosity function) of the loops, there are no other significant differences
between models computed with different EO scales.
The  SFR, estimated from the best-match model in each case, increases by less than $\sim$15\% at increasing EO mixing,
and so it is not significantly affected by the choice of this parameter. We also find that a Salpeter slope for intermediate  and massive stars
is, in general, the preferred choice to reproduce the observed luminosity function.

It is worth noticing that the
mixing scales required to
reproduce the observed loops, EO=2\HP or EO=4\HP, are definitely larger than the maximum values compatible with e.g. the location of the RGB bump in evolved low mass stars, which can be well reproduced by an envelope overshooting
not exceeding EO=0.7\HP,  the value adopted in {\sl\,PARSEC}~V1.1 for
intermediate mass stars.
Thus the results suggest a strong dependence of the mixing scale below
the formal Schwarzschild border of the outer envelope, with the mass or luminosity of the star.
A possible mechanism at the origin of such enhanced mixing could be the large discontinuity in angular momentum that develops between
the outer envelope and the inner core during the first dredge-UP, when stellar rotation is included. The shear generated by this discontinuity could give rise to the required enhancement.
However it is interesting to note that current models that consistently include rotational mixing in the whole star interiors, do not show  larger loops \citep{heger_langer_00, Georgy_etal13},
and thus would face the same difficulty.
More extended loops could be possibly obtained by adopting
the Ledoux criterion instead of the Schwarzschild one,
because the former is known to favour the development of loops,
at least in the domain of massive stars \citep{chiosi_summa_70}.
Though disfavored by the
analisys of \citet{Kato66},
the Ledoux criterion has recently received some observational support \citep{Georgy14},
and its consequences are worth being explored in a systematic way.
All this renders the problem of the blue loops and
 the origin of such enhanced mixing quite challenging.
At the moment our simulations support the possibility that, in the framework
of a mild efficiency of convective core overshooting, the mixing by the first dredge-UP becomes more efficient at increasing luminosity, and could
reach a few pressure scale heights below the formal Schwarzschild border, in the domain of intermediate and massive stars.
In the future we will
perform the same comparison with  galaxies of different metallicity, a work that is already in progress. Furthermore it should be worth to test models that use different schemes, such as rotational mixing and/or  diffusive approach of overshooting, to see if there are other mixing schemes that are able to produce internal chemical H profiles that allow the development of more extend loops.

\section{Acknowledgements}
Based on observations with the Hubble Space Telescope, program HST-GO 11079 and 10915.
We thank M. Barbieri, Z. Y. Cai, S. Charlot,  Y. Chen, A. Grotsch-Noels, M. Mapelli, M. Spera and S. Zaggia for helpful discussions.
We thank the anonymous referee for useful suggestions.
P.M. acknowledges support from {\em Progetto di Ateneo 2012}, University of Padova,  ID: CPDA125588/12.

\bibliographystyle{mn2e/mn2e_new} 
\bibliography{dwarfs} 
%
%
\label{lastpage}
\end{document}